\documentclass[
  reprint,
  aps,
  prx,
  superscriptaddress,
  nofootinbib
]{revtex4-2}

\usepackage{graphicx}
\usepackage{gensymb}
\usepackage{mhchem}
\usepackage{xcolor}
\usepackage{amsmath,amssymb}
\usepackage{comment}
\usepackage[caption=false]{subfig}
\usepackage{makecell}
\usepackage{verbatim}
\usepackage{hyperref}
\usepackage{array}
\usepackage{placeins}
\begin{document}

\title{Compact narrowband photon-pair generation by slow-light spectral engineering}
\author{Ashwith Prabhu}
\affiliation{Department of Physics, University of Illinois Urbana-Champaign, Urbana, IL 61801 USA}

\author{Elizabeth A. Goldschmidt}
\altaffiliation{goldschm@illinois.edu}
\affiliation{Department of Physics, University of Illinois Urbana-Champaign, Urbana, IL 61801 USA}

\begin{abstract}
Efficiently generating photon pairs with high heralding efficiency and high single photon purity that are bandwidth matched to quantum emitters, quantum memories, and other matter-based qubits is critical for quantum networking applications. However, nonlinear optics-based sources require substantial spectral engineering to overcome the orders of magnitude bandwidth mismatch between those sources and qubit systems. A popular solution is cavity-enhanced spontaneous parametric down conversion (SPDC) where the cavity sets the photon bandwidth and simultaneously enhances the spectral brightness of the SPDC. Bulk, free-space configurations are generally required to achieve the MHz-scale bandwidths required to interface with most qubit systems. Replicating these in scalable integrated photonic architectures is an ongoing challenge due to the much higher propagation losses that limit the size and linewidth of chip-based resonators. We show here how an intra-cavity slow light medium, acting as an ultra-narrow filter, would enable narrowband photon pair generation in broadband cavities with high single photon purity and without compromising the heralding efficiency. We show that such metrics can be readily realized in erbium doped thin-film lithium niobate microrings using realistic design parameters.      
\end{abstract}

\maketitle

\section{Introduction}
Photon pairs generated via spontaneous parametric down conversion (SPDC) and other nonlinear processes are a vital workhorse for quantum networking and other applications \cite{couteau2018spontaneous, kwiat1995new}. One feature of such sources is the ability to control the spectrum of the emitted photons by engineering the phase-matching function through temperature, angle, periodic poling, and many other methods\,\cite{byer1997quasi, burlakov2001collinear, vodopyanov2004optical, gong2011compact,tanzilli2002ppln,kuklewicz2004high, kim2006phase}. However, the bandwidth of the phase-matching function is typically many THz or more, far too large for  applications that require the photons to interface with quantum memories or other matter-based qubit platforms where MHz-scale bandwidths are typical. This bandwidth mismatch is typically overcome by generating the photon pairs in a high-finesse optical cavity whose linewidth is engineered to match the desired photon bandwidth\,\cite{luo2015direct, rielander2016cavity, slattery2015narrow, mataji2023narrow}. The cavity plays two important roles, in addition to narrowing the bandwidth of the produced photon pairs, it also enhances the nonlinear process within that narrow bandwidth to increase the spectral brightness (rate of photons produced per bandwidth)\,\cite{ou1999cavity, scholz2009analytical}. Such schemes have been used to demonstrate entanglement distribution between quantum memories in a range of systems \cite{sangouard2011quantum, zhang2011preparation, rielander2014quantum, clausen2011quantum, saglamyurek2011broadband, schunk2015interfacing, lenhard2015telecom, simon2007quantum}. 

In order to build quantum networks and other devices at scale, photon pair production should be implemented in scalable integrated photonic platforms. However, nanophotonic devices suffer from propagation losses that limit typical resonator bandwidths to $\gtrsim$~GHz, thus limiting chip-scale photon pair production to the GHz-regime\,\cite{guo2017parametric, ma2020ultrabright, luo2017chip, zhao2022ingap, li2025down, lu2019periodically}. Achieving MHz-bandwidth photons is accomplished with bulk, free-space cavities that have a large footprint and are not particularly robust or scalable. Our breakthrough is the recognition that an ultra-narrow intra-cavity absorptive filter can act as a slow light medium and extend the effective length of a compact cavity without adding any round trip loss\,\cite{barya2025ultra, sabooni2013spectral}. While we focus here on implementing this via a process like spectral hole burning or electromagnetically induced transparency in an optically dense ensemble of coherent emitters, any method of introducing a filter into the resonator mode would have the same effect. 

Thus, we propose a scheme that would be physically inaccessible without the slow light medium where the effective resonator length, and group velocity, are orders of magnitude different for the different fields involved in the pair production (pump, signal, and idler). We consider the doubly filtered case where both signal and idler are in narrowed modes and the singly filtered case where only the signal is in a narrowed mode with the idler in a ``bare" mode with similar bandwidth and group velocity as the pump. We investigate here, analytically and numerically, the effect on the bandwidth, photon pair generation rate, heralding efficiency, and spectral purity for photon pairs produced via SPDC in a cavity with an ultra-narrow intra-cavity filter acting on the signal and/or idler modes. We find that for experimentally accessible parameters, photons with MHz-level bandwidths could be produced in nanophotonic devices with propagation losses that limit ``bare" resonator linewidths to the GHz scale. 

\begin{figure*} 
    \centering
    \subfloat[]{%
        \includegraphics[width=0.32\textwidth]{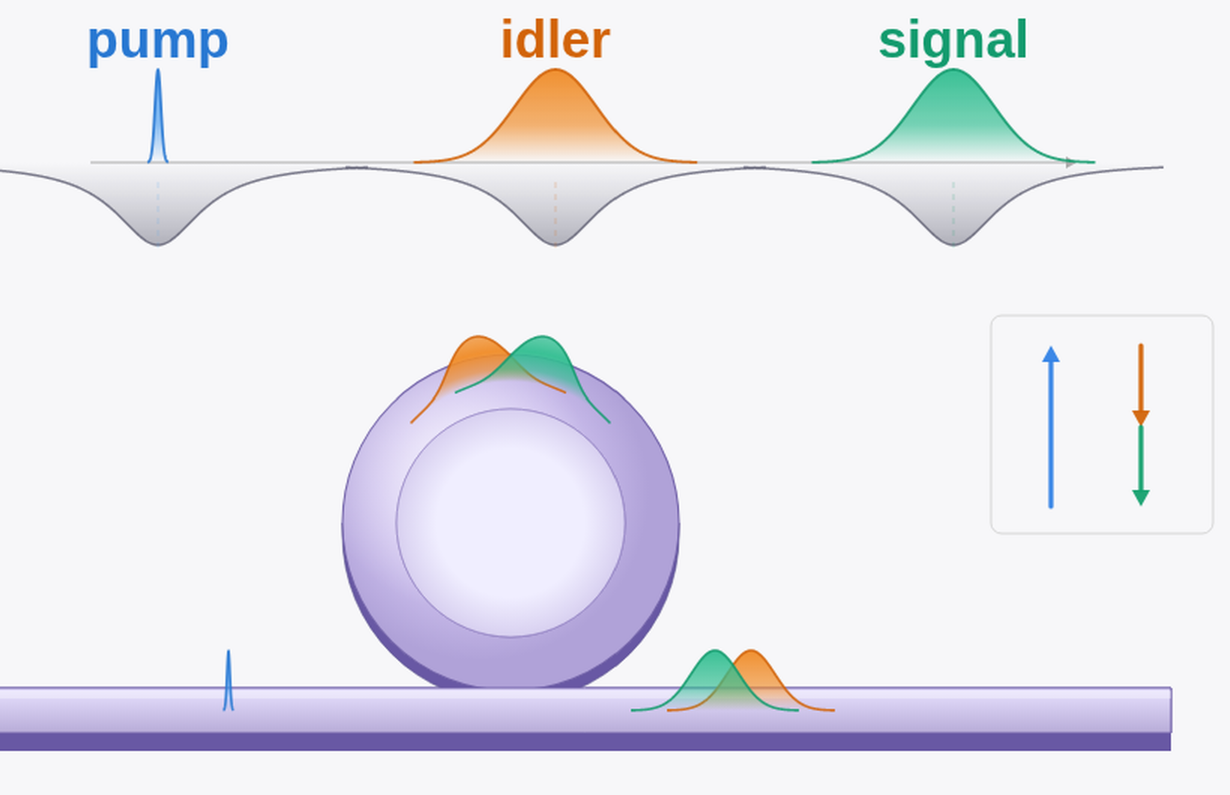}%
        \label{fig:schematicspdcpanel1}%
    }\hfill
     \subfloat[]{%
        \includegraphics[width=0.32\textwidth]{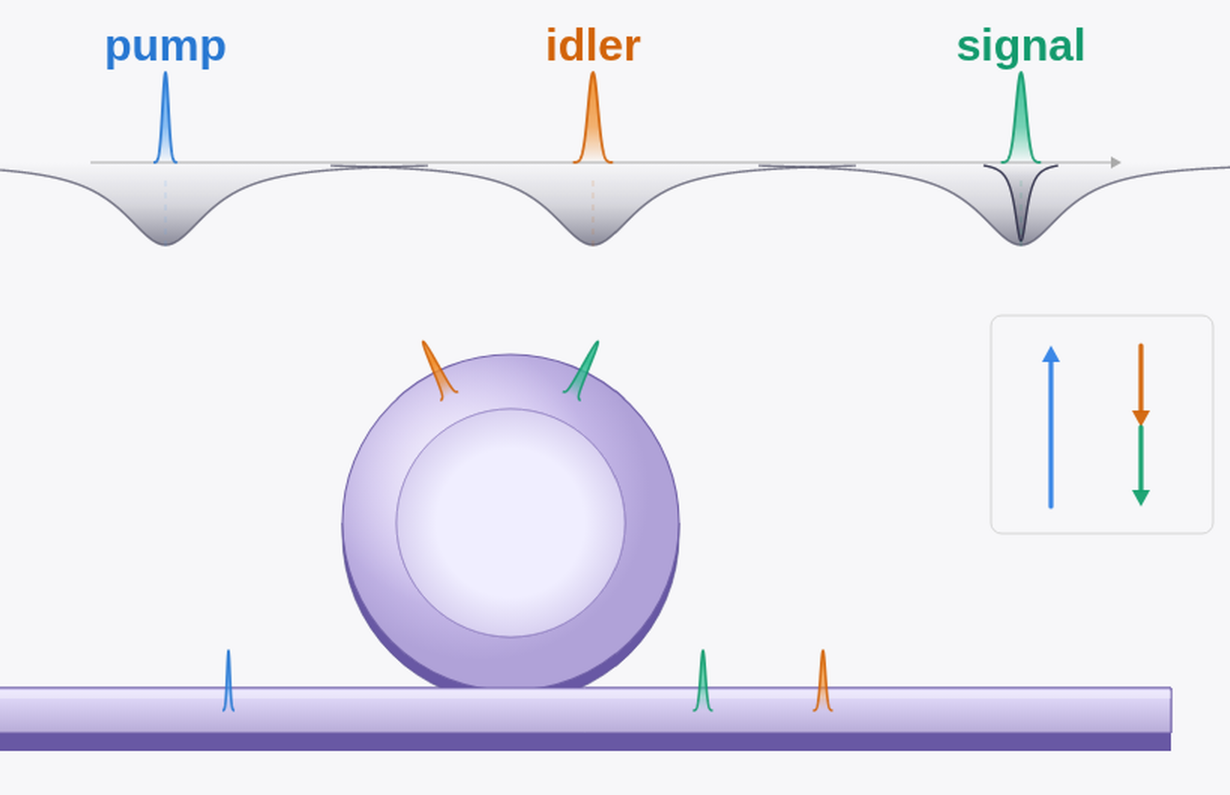}%
        \label{fig:schematicspdcpanel2}%
    }\hfill
     \subfloat[]{%
        \includegraphics[width=0.32\textwidth]{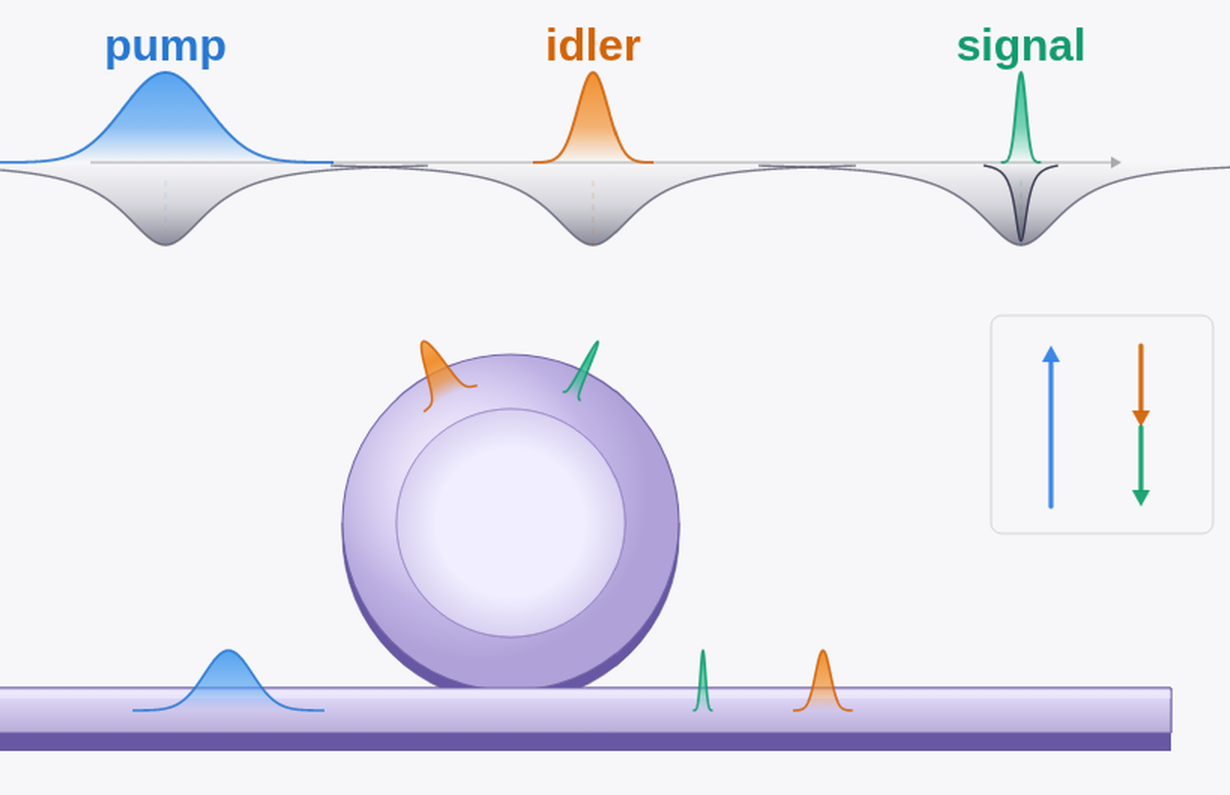}%
        \label{fig:schematicspdcpanel3}%
    }
    \caption{ \textbf{Schematic for SPDC in different regimes. For the bare cavity (a) a narrowband pump generates photons whose bandwidths match their respective cavity modes. For a cavity with a narrowed signal mode (b) a narrowband pump generates photons both whose bandwidths match the \textit{signal} mode bandwidth, and (c) a broader pump generates a signal photon with bandwidth that equals its cavity mode and an idler photon with tunable bandwidth. Note: all pulse widths shown indicate the \textit{spectral} width of the pulse.}}
    \label{fig:schematicspdc}
\end{figure*}

We focus on the applications to on-chip devices, particularly the parameter regime that can be realized in an erbium-doped thin film lithium niobate microring resonator \cite{barya2025ultra}, and thus assume throughout a triply resonant cavity where pump, signal, and idler are all resonant with bare cavity modes and suitable phase-matching has been ensured via periodic poling or some other method. We note that the scheme proposed here could also be used to reduce the footprint of bulk configurations and all results apply equally to such a configuration \cite{fekete2013ultranarrow}. We first consider continuous wave pumping which allows us to derive analytic expressions for the relevant metrics and qualitatively compare the bare cavity case against the filtered cavity. Doubly and singly filtered SPDC can lead to qualitatively different outcomes under the action of a continuous wave pump. We treat the two cases separately in sections II \& III. In section IV, we investigate the effect of using a quasi-continuous pump with a bandwidth that is tunable over the relevant range, with an emphasis on the spectral purity and heralding efficiency of the generated photons.  

\begin{table*}[tb]
\begin{ruledtabular}
\begin{tabular}{cccc}
\\[-12ex]
\textbf{} & \textbf{Bare Cavity} & \textbf{Filtered Cavity}\\
\hline 
\\[-10ex]
\textbf{Signal/Idler Bandwidth ($\Delta \omega$)} & $0.64\kappa_a$ & $0.64\kappa_a\frac{1}{n_g}$  \\[-3ex]
\textbf{Pair Generation Rate ($R$)} & $\frac{4g^2|\beta|^2}{\kappa_a}$ &  $\frac{4g^2|\beta|^2}{\kappa_a}\frac{1}{n_g}$ \\[-3ex]
\textbf{Spectral Brightness\,($\frac{R}{\Delta\omega}$)} & $\frac{6.25g^2|\beta|^2}{\kappa_a^2}$ & $\frac{6.25g^2|\beta|^2}{\kappa_a^2}$\\[-3ex]
\textbf{Correlation Time\,($\tau_c$)} & $\frac{1}{\kappa_a}$ & $\frac{1}{\kappa_a}n_g$\\[-3ex]
\textbf{Second Order Correlation function ($g^{(2)}(\tau)$)} & $1 +  \frac{1}{4R\tau_c}e^{-\frac{|\tau|}{\tau_c}}$ & $1 +  \frac{1}{4R\tau_c}e^{-\frac{|\tau|}{\tau_c}}$\\[-3ex]
\textbf{Heralding efficiency ($\eta^h$)} & $\frac{{\kappa_a}_{\text{ext}}}{\kappa_a}\left(1-e^{-\frac{T}{2\tau_c}}\right)$ & $\frac{{\kappa_a}_{\text{ext}}}{\kappa_a}\left(1-e^{-\frac{T}{2\tau_c}}\right)$
\end{tabular}
\end{ruledtabular}
\label{table:degenSPDCmetrics}
\caption{\textbf{Degenerate SPDC metrics for bare cavity and cavity with an extended filter. Bandwidth and pair generation rate are both reduced by a factor of $n_g$, leaving spectral brightness unchanged, while the correlation time is increased by a factor of $n_g$, spreading the correlations over a longer time without reducing the peak heralding efficiency.}}
\end{table*}

\section{Doubly Filtered pair production}
Photon pairs produced via SPDC must obey energy conservation and phase matching. The former requires that $\omega_p=\omega_s+\omega_i$, where the subscripts $p,s,i$ stand for pump, signal, and idler, respectively, and $\omega$ denotes the angular frequency of that field. In this section and the following one, we assume the pump is a continuous wave field with negligible bandwidth (the effect of tuning the pump bandwidth is investigated in section IV). Phase-matching for SPDC can be engineered to occur at nearly any arbitrary set of wavelengths where the  medium supports nonlinear processes and energy conservation holds, with typical phase-matching bandwidths in the THz regime \cite{boyd2008nonlinear}. Putting the nonlinear medium in a cavity enhances pair production in the cavity spectral modes and suppresses it in the spectral regions between those modes \cite{ou1999cavity}. There is extensive work in the literature describing such cavity-enhanced SPDC for the three cases: the singly-resonant case where just one of the signal or idler fields is resonant with the cavity, the doubly-resonant case where both signal and idler are resonant with cavity modes, and the triply-resonant case where pump, signal, and idler are all resonant with cavity modes \cite{jeronimo2010theory, moqanaki2019novel, scholz2007narrow}. For nanophotonic devices where ring resonators are commonly used, only the triply-resonant case is achievable, and thus we focus here only on the case where all three fields are resonant with bare cavity modes before we introduce any intra-cavity filter. 

In this section, we investigate the case that both signal and idler fields see narrow intra-cavity absorptive filters. We assume throughout the text that the band-pass window overlaps with only one cavity mode while suppressing the neighboring modes. We first focus on the case where the signal and idler are degenerate and subjected to a single extended intra-cavity filter. We follow standard treatments for cavity-enhanced SPDC to derive the temporal and spectral properties of photons generated in this regime, and then use those quantities to calculate the photons' rate, correlation, and purity. This is highly instructive and brings forth the essential features of the intra-cavity filtering effect.
 
Under the appropriate phase matching conditions, a continuous wave pump excites a cavity mode, $c$, resonant at frequency $\omega_c$ and degenerate photon pairs are emitted in another cavity mode, $a$, at frequency $\omega_a$ with bare linewidth $\kappa_a$. The Hamiltonian describing degenerate SPDC in a triply resonant cavity is as follows. 
\begin{align}
H=&\omega_a a^\dagger a +\omega_c c^\dagger c + g\left((a^\dagger)^2c+a^2c^\dagger\right)\nonumber\\
&+i\epsilon_p(c^\dagger e^{-i\omega_pt}-ce^{i\omega_pt})
\end{align}
Here, $g$ is the nonlinear coupling strength and we have set $\hbar=1$. The quantity $\epsilon_p=\sqrt{{\kappa_c}_{\text{ext}}\frac{P_p}{\hbar\omega_p}}$ is the effective amplitude of a pump with power $P_p$ coupled to the cavity mode $c$ via the external coupling coefficient ${\kappa_c}_{\text{ext}}$. Under the assumption that the pump remains undepleted, and in the frame of reference of half the pump drive frequency, $\omega_p/2$, the following equation describes the dynamics of the degenerate signal and idler cavity mode, $a$.
\begin{align}
\dot{a}=-i\delta_a a -2ig\beta a^\dagger -\frac{\kappa_a}{2}a+\sqrt{{\kappa_a}} a_{\text{in}}
\end{align}
Here, $\beta$ is the steady state amplitude of the pump cavity mode, detuned from the pump drive by $\delta_c\equiv \omega_c-\omega_p$ such that $\beta\equiv \frac{\epsilon_p}{\frac{\kappa_c}{2}+i\delta_c}$. The vacuum mode, $a_{\text{in}}$ couples to the cavity mode via the coupling coefficient $\sqrt{\kappa_a}$\,\cite{walls2008quantum}. The cavity mode decays with rate $\kappa_a$ and is detuned from the pump drive half frequency by $\delta_a\equiv \omega_a-\frac{\omega_p}{2}$. We now introduce a dissipative filter with a band-pass full width half-maximum of $\frac{\Delta}{2\pi}\ll\kappa_a$. We assume a Lorentzian filter shape with zero loss at the central frequency and loss at a rate $\kappa_{\text{abs}}$ far from the filter center. The effect of the filter on the cavity dynamics is captured as follows:
\begin{align}
    \dot{a}=&-i\delta_a a -2ig\beta a^\dagger -\frac{\kappa_a}{2}a-\frac{\kappa_{\text{abs}}}{2}a\nonumber\\
    &+\frac{\kappa_{\text{abs}}}{2}\frac{\Delta}{2}\int_{0}^td\tau\, e^{-(\frac{\Delta}{2}+i\delta_a)(t-\tau)} a(\tau)+\sqrt{{\kappa_a}} a_{\text{in}}
\end{align}
Phenomenologically, the band-pass filter creates a sharp dispersion profile within its spectral window, leading to a large group index, $n_g=\frac{\kappa_a+\kappa_{\text{abs}}}{\Delta}\gg1$, and a slow group velocity $v_g\approx c/n_g$, thereby, extending both the cavity round trip time and ringdown time by a factor of $n_g$ \cite{barya2025ultra}. As the filter does not add any round trip loss for a resonant field, this effectively makes the cavity appear to be \textit{longer} by a factor of $n_g$. Importantly, this effective lengthening description correctly captures the narrowing of the cavity mode without any increase in the circulating electric field.   

This notion of cavity mode narrowing can be formalized by a Fano-Feshbach type projection onto a low-energy subspace of the Hamiltonian\,\cite{jing2012feshbach, chruscinski2013feshbach, fano1961effects, feshbach1958unified}. Following the projection, the Fourier domain equation governing the dynamics of the narrowed cavity mode is:
\begin{widetext}
\begin{align}
    -i\omega\tilde{a}(\omega)=-\frac{\kappa_a}{2}\frac{\Delta}{\kappa_a+\kappa_{\text{abs}}}\tilde{a}(\omega)-2ig\beta \frac{\Delta}{\kappa_a+\kappa_{\text{abs}}}\tilde{a}^\dagger(-\omega-2\delta_a)
    +\sqrt{{\kappa_a}}\frac{\Delta}{\kappa_a+\kappa_{\text{abs}}}a_{\text{in}}(\omega)
\end{align}
\end{widetext}
Here, $\tilde{a}$ denotes the narrowed cavity mode whose full-width half maximum, $\kappa_a\frac{\Delta}{\kappa_a+\kappa_{\text{abs}}}=n_g \kappa_a$, is reduced from the bare width by a factor equal to the group index of refraction at the center of the filter\,\cite{barya2025ultra}. Notably, the narrowed cavity width is much narrower than the filter itself ($ \kappa_a/n_g\ll\Delta$) as long as $\kappa_{\rm{abs}}\gg\kappa_a$. We calculate the the spectral density,\,$S_a(\omega, \omega^\prime)$, of the generated photons to extract their bandwidth, $\Delta\omega$, and generation rate, $R$\,\cite{ou1999cavity}. 
\begin{align}
&S_a(\omega, \omega^\prime)=\kappa_a\langle \tilde{a}^\dagger(\omega) \tilde{a}(\omega^\prime)\rangle \nonumber\\
&\Delta\omega=0.64\kappa_a/n_g\\
&R=\frac{1}{2}\int_{-\infty}^{\infty}\frac{d\omega}{2\pi}\frac{d\omega^\prime}{2\pi}S_a(\omega, \omega^\prime) e^{i(\omega-\omega^\prime)t}\approx\frac{4g^2|\beta|^2}{n_g\kappa_a}
\end{align}

We see that a direct consequence of the cavity mode narrowing is a decrease in both the signal/idler bandwidth and the generation rate by a factor of $\approx n_g$. Crucially, this identical scaling means that the photon pair generation rate per bandwidth, or \textit{spectral brightness}, is unchanged from the unfiltered scheme. Because the photon bandwidth is approximately $0.64\kappa_a/n_g$ rather than the filter width, $\Delta$, we see that this method generates photons that are not only much narrower than the bare cavity, but also much narrower than the filter itself. This is important given practical limits on how narrow of a filter can be implemented in many regimes. We will put realistic values to all these quantities in section \ref{sec:broadband}.

So far, one would obtain similar results with a post-cavity filter, noting that a post-cavity filter would have to be narrower to achieve the same photon bandwidth. However, we will see in \ref{sec:broadband} that a post-cavity filter leads to an unavoidable tradeoff between spectral purity and heralding efficiency as we move away from exclusively continuous wave pump fields\,\cite{meyer2017limits}. While this tradeoff is not relevant for the  continuous-wave pumps, which always lead to large spectral entanglement between the photons, but we can obtain an analytical expression for the heralding efficiency in this case, which is illustrative. The heralding efficiency is the probability of detecting one photon within some time window, $T$, of detecting the other and is derived from the normalized second order cross-correlation between signal and idler, $g^{(2)}(\tau)$.

\begin{align}
    &g^{(2)}(\tau)=\frac{\langle \tilde{a}^\dagger_{\text{out}}(t)\tilde{a}^\dagger_{\text{out}}(t+\tau)\tilde{a}_{\text{out}}(t+\tau)\tilde{a}_{\text{out}}(t)\rangle}{\langle \tilde{a}^\dagger_{\text{out}}(t)\tilde{a}_{\text{out}}(t)\rangle\langle\tilde{a}^\dagger_{\text{out}}(t+\tau)\tilde{a}_{\text{out}}(t+\tau))\rangle}\nonumber\\
    &\eta^h=\frac{2{\kappa_a}_{\text{ext}}}{\kappa_a}R\int_{-T/2}^{T/2}d\tau\,\left(g^{(2)}(\tau)-1\right)
\end{align}
 The output field is coupled to the input vacuum field and the intra-cavity field via the usual input-output relations\,\cite{walls2008quantum}. Note that the second order correlation function is independent of time, $t$. This is strictly true only for a continuous wave pump.  We also note that throughout this work we assume weak pumping and stay far from the regime of optical parametric oscillation. The effect of the narrowing on these metrics have been compiled in Table \ref{table:degenSPDCmetrics}. To allow a convenient side-by-side comparison between the bare and narrowed cavity, we have set the detuning, $\delta_a$, to be zero. The expressions for the most general case and the algebraic details involved in attaining them have been relegated to the supplemental material\,\cite{SuppMat}. Taking into account all the metrics, it is evident that the intra-cavity filter can be fully thought of as increasing the effective length of the cavity for the signal and idler modes. We see in the following sections that $n_g$ as large as $\sim1000$ can be achieved assuming reasonable parameters, meaning a dramatically smaller footprint can be used for generating narrowband photons using this approach.

 We note here that the case of doubly-filtered \textit{non-degenerate} SPDC exhibits the same narrowing compared to the bare cavity that we just described for the doubly-filtered degenerate SPDC. One difference is that any phase mismatch induced by the different group indices arising from non-identical intra-cavity filters must be accounted for in the non-degenerate case. The extreme end of this asymmetry is the single-filtered case where only one photon is subject to a filter narrower than the bare resonator. This case is analyzed in detail in the next section.

\section{Singly Filtered Pair Production}
We saw that the doubly-filtered case makes the cavity appear like a longer and narrower cavity for the signal and idler photons than it is in the absence of the filter. The case where only one of the photons is produced in a filtered mode has some new and unique features, as we see in this section and the following section.  

For the singly filtered case, the signal and idler photons are emitted into two spectrally disparate cavity modes, $a$ and $b$, resonant at frequencies, $\omega_a$ and $\omega_b$ respectively. Under similar pumping conditions as the degenerate case and assuming the phase-matching is appropriately engineered, the Hamiltonian is:
\begin{align}
H=&\omega_a a^\dagger a + \omega_b b^\dagger b + \omega_cc^\dagger c + 2g\left(a^\dagger c^\dagger b + acb^\dagger\right) \nonumber\\
&+ i\epsilon_p(c^\dagger e^{-i\omega_pt}-ce^{i\omega_pt}) 
\end{align}
The dynamics of the $a$ and $b$ cavity modes follows from the above Hamiltonian. We introduce a Lorentz dissipative filter with its spectral window centered at $\omega_a$ and assume that the $b$ cavity mode is spectrally so far removed that it is unaffected by the filter. In the Fourier domain, following the Fano-Feshbach type projection, we have:
\begin{widetext}
\begin{align}
-i\omega\tilde{a}(\omega)=-\frac{\kappa_a}{2}\frac{1}{n_g}\tilde{a}(\omega)-2ig\beta F(\omega) \frac{1}{n_g}b^\dagger(-\omega-2\delta_a)
+\sqrt{{\kappa_a}}\frac{1}{n_g}a_{\text{in}}(\omega)\\
 -i\omega b^\dagger(-\omega-2\delta_a)= i(\delta_b+\delta_a) b^\dagger (-\omega-2\delta_a)
-\frac{\kappa_c}{2} b^\dagger(-\omega-2\delta_a)+2ig\beta^*F^*(\omega)\tilde{a}(\omega)+\sqrt{{\kappa_b}}b_{\text{in}}(\omega)
\end{align}
\end{widetext}
We have chosen to operate in the frame of reference of resonance frequency of cavity mode $a$ for convenience. In the above set of equations, $\delta_b\equiv \omega_b-\omega_p/2$  is the detuning of cavity mode $b$ from the pump drive half frequency.  It is coupled to vacuum mode $b_{\text{in}}(\omega)$ and decays with total rate $\kappa_b$. In the doubly filtered case, the additional dispersion of the intra-cavity filter did not impact the phase matching because both signal and idler saw identical dispersion. Here we must consider the effect of the accumulated phase mismatch due to the different group velocities. The phase matching function, $F(\omega)$, captures the phase mismatch over one round trip between the $\tilde{a}$ and $b$ modes. We assume here, and throughout this work, that the ``bare" phase-matching in the absence of the intra-cavity filter is perfect, including the effects of material or waveguide dispersion, in order to elucidate the impact of the filter, specifically. We also assume from here on that the bare cavity modes have the same bandwidth ($\kappa_a=\kappa_b=\kappa$ and $\kappa_{a_{\rm{ext}}}=\kappa_{a_{\rm{ext}}}=\kappa_{\rm{ext}}$).
\begin{align}
    F(\omega)&=\text{sinc}\left(\frac{\omega\tau_{\text{rt}}}{2}\right)\\
    \tau_{\text{rt}}&=\frac{nL}{c}(n_g-1)\approx \frac{nL}{c}n_g 
\end{align}
$L$ and $n$ denote the round trip length of the cavity and the effective refractive index of the bare cavity mode. $\tau_{\text{rt}}$ is the difference in the round trip time between the two modes, which is approximately equal to the much longer round trip time for the narrowed mode. The intrinsic spectral density for signal or idler photons has the following general form for a continuous pump:
\begin{align}
    S_a(\omega, \omega^\prime)=S_b(\omega, \omega^\prime)=C(\omega, \omega^\prime)|F(\omega)|^2\delta(\omega-\omega^\prime)
\end{align}
The term, $C(\omega, \omega^\prime)$ is the joint cavity response function, which is simply the product of the individual response functions under our weak pumping assumption. With no intra-cavity filter, the phase matching function is unity and the bandwidth of the signal and idler photons is proportional to the cavity width. In the presence of an intra-cavity filter, the cavity widths of the two modes become highly asymmetric and the width of the joint cavity response function is simply the narrower cavity width, $\kappa/n_g$. The phase-matching function, in fact, remains unity in the relevant spectral region as long as the cavity finesse, $\mathcal{F}$, is larger than 2.26, a condition that is met in virtually any realistic scenario. Thus, the result of narrowing the mode for one of the photons is to narrow the bandwidth of both photons, preserving the spectral brightness achieved in the bare cavity.  
 \begin{figure}
\centering
 \includegraphics[width=\linewidth]{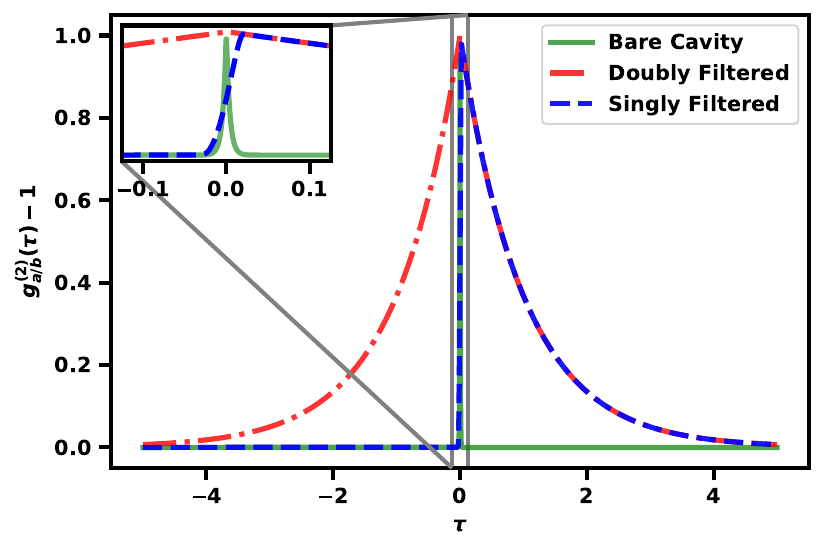}
 \caption{\textbf{Normalized second order cross-correlation function as a function of the time different, $\tau$, between detecting the idler photon and the signal photon. The value of  $g^{(2)}_{a/b}-1$ is shown normalized to the peak value of $(\kappa/4g|\beta|)^2$, which is the same for all curves. The time axes is in units of the \textit{filtered} correlation time $\tau_c=n_g/\kappa$.}}
\label{fig:g2comparison}
\end{figure}

\begin{table*}[t]
\begin{ruledtabular}
\begin{tabular}{ccc}
\\[-12ex]
\textbf{Metric} & \textbf{Bare Cavity} & \textbf{Filtered Cavity}\\
\hline
\\[-10ex]
\textbf{$\Delta \omega$} & $0.64\kappa$ & $\frac{\kappa}{n_g}$  \\
\textbf{$R$} & $\frac{8g^2|\beta|^2}{\kappa}$ &  $\frac{16g^2|\beta|^2}{\kappa}\frac{1}{n_g}$ \\
\textbf{$\frac{R}{\Delta\omega}$} & $\frac{12.5g^2|\beta|^2}{\kappa^2}$ & $\frac{16g^2|\beta|^2}{\kappa^2}$\\
\textbf{$g^{(2)}_{a/b}(\tau)$} & $1+\frac{\kappa^2}{16g^2|\beta|^2}e^{-\kappa|\tau|}$
&$\begin{aligned}[t]
\tau < -\frac{\tau_{\text{rt}}}{2}: \quad &
1+\frac{\kappa^2}{16g^2|\beta|^2}
\left[
e^{\kappa\tau}\,
\mathrm{sinc}^2\!\left(i\frac{\kappa\tau_{\text{rt}}}{4}\right)
\right]
\\[8pt]

|\tau| \le \frac{\tau_{\text{rt}}}{2}: \quad &
1+\frac{\kappa^2}{16g^2|\beta|^2}\frac{4}{\kappa^2\tau_{\text{rt}}^2}
\left[
e^{\frac{\kappa}{2}\left(\tau-\frac{\tau_{\text{rt}}}{2}\right)}
+n_ge^{-\frac{\kappa}{2n_g}\left(\tau+\frac{\tau_{\text{rt}}}{2}\right)}
-1-n_g
\right]^2
\\[10pt]

\tau > \frac{\tau_{\text{rt}}}{2}: \quad &
1+\frac{\kappa^2}{16g^2|\beta|^2}
\left[
e^{-\frac{\kappa}{n_g}\tau}\,
\mathrm{sinc}^2\!\left(i\frac{\kappa\tau_{\text{rt}}}{4n_g}\right)
\right]
\end{aligned}$\\
$\eta^h$ & $\frac{\kappa_{\text{ext}}}{\kappa}(1-e^{-\kappa\frac{T}{2}})$ &   $\quad \frac{\kappa_{\text{ext}}}{\kappa}\left(1-e^{-\frac{\kappa}{n_g}\frac{T}{2}}\right)-\frac{\kappa_{\text{ext}}}{\kappa}\frac{\kappa\tau_{\text{rt}}}{6n_g}$ 
\end{tabular}
\end{ruledtabular}
\label{table:nondegenSPDCmetrics}
\caption{\textbf{Non-degenerate SPDC metrics for bare cavity and cavity with a single extended filter. The metrics from top to bottom are the bandwidth, intrinsic pair generation rate, spectral brightness, normalized  second order correlation function and heralding efficiency. Note that the expression for the heralding efficiency assumes $T>\tau_{\text{rt}}$, the full expression can be found in the supplemental materials \cite{SuppMat}.}}
\end{table*}

 We now turn our attention to the temporal correlation between the two modes. The normalized second order correlation function conditioned on the $b$ mode, $g^{(2)}_{a/b}(\tau)$ and the heralding efficiency $\eta^h$ have the following form:

 \begin{align}
 g^{(2)}_{a/b}(\tau)=&\frac{\langle b_{\text{out}}^\dagger(t)\tilde{a}_\text{out}^\dagger(t+\tau)\tilde{a}_{\text{out}}(t+\tau)b_{\text{out}}(t)\rangle}{\langle b_{\text{out}}^\dagger(t)b_{\text{out}}(t)\rangle\langle\tilde{a}_{\text{out}}^\dagger(t+\tau)\tilde{a}_{\text{out}}(t+\tau)\rangle}\\
\eta^h=&\frac{{\kappa_b}_\text{ext}}{\kappa_b}R\int_{-T/2}^0d\tau\left(g^{(2)}_{a/b}(\tau)-1\right)\nonumber\\
&+\frac{{\kappa_a}_{\text{ext}}}{\kappa_a}R\int_0^{T/2}d\tau\left(g^{(2)}_{a/b}(\tau)-1\right)
 \end{align}
 
 We illustrate the effect of intra-cavity filtering in  Fig\,\ref{fig:g2comparison} by plotting $g^{(2)}(\tau)$ for the bare cavity, the doubly-filtered case, and the singly-filtered case in a typical parameter regime. We see that the correlation function decays at the ``bare" rate $\kappa$ for the ``slower,"  a-mode photon arriving first and at the narrowed rate of $\kappa/n_g$ for the more natural case of the``faster", b-mode photon arriving first. Here, slower and faster are referring both to the group velocity of the photon in the resonator and the rate at which the photon escapes its resonator mode. Loosely speaking, the correlation function indicates the probability of detecting the ``slow," signal photon at a time $\tau$ following the detection of the ``fast," idler photon. The asymmetry of the function and the extended correlation time relative to the bare cavity case relates to the increased time taken by the signal photons occupying the filtered $a$ mode to escape the cavity. For sufficiently long detection window, the maximum achievable heralding efficiency, $\eta^h$ is identical to the bare cavity case up to a small first order correction introduced by the phase matching function. We have compiled the metrics for the bare and singly filtered cavity case in Table \ref{table:nondegenSPDCmetrics}. We have set the two photon detuning to zero $(\delta_a+\delta_b=0)$ and assumed that the bare cavity modes have the same width, $\kappa$. The expressions for the more general case can be found in the supplemental material\,\cite{SuppMat}.

 \begin{figure*}
    \centering
    \subfloat[]{\includegraphics[width=0.3\textwidth]{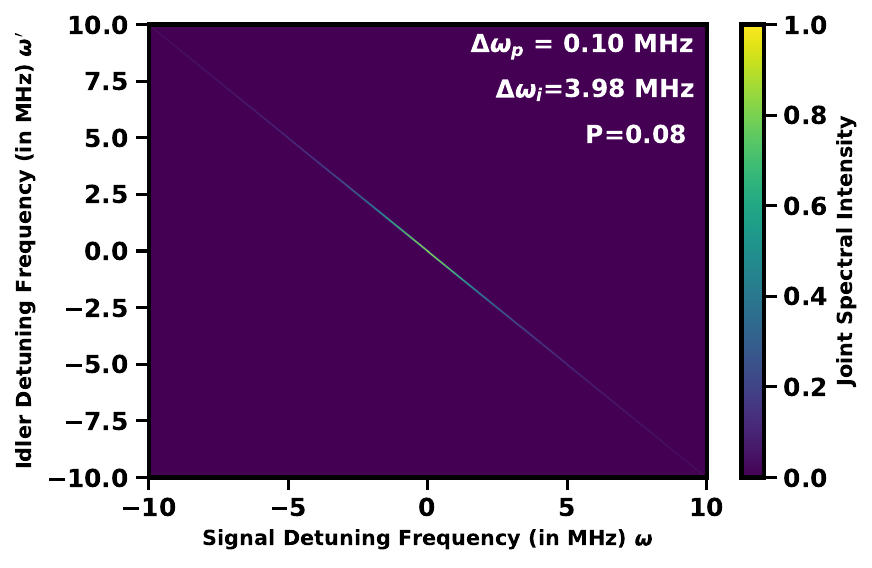}}
    \hfill
    \subfloat[]{\includegraphics[width=0.3\textwidth]{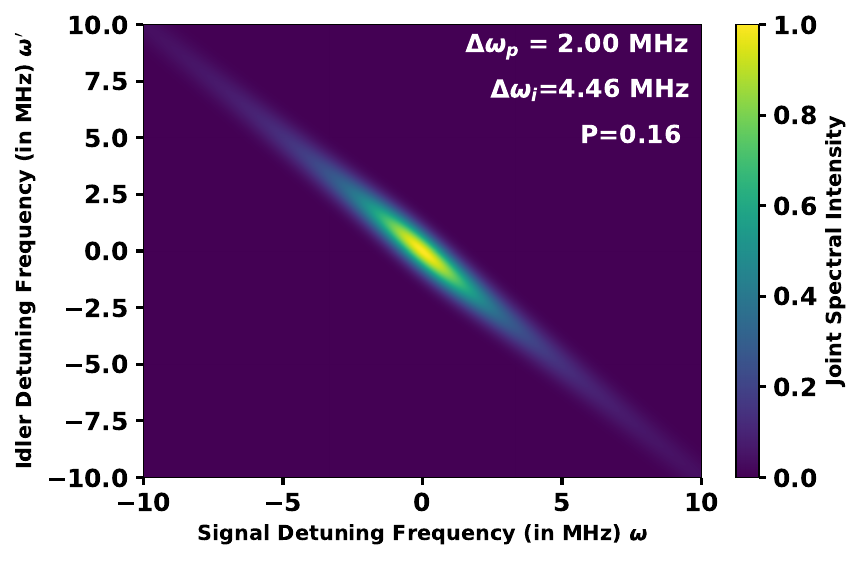}}
    \hfill
    \subfloat[]{\includegraphics[width=0.3\textwidth]{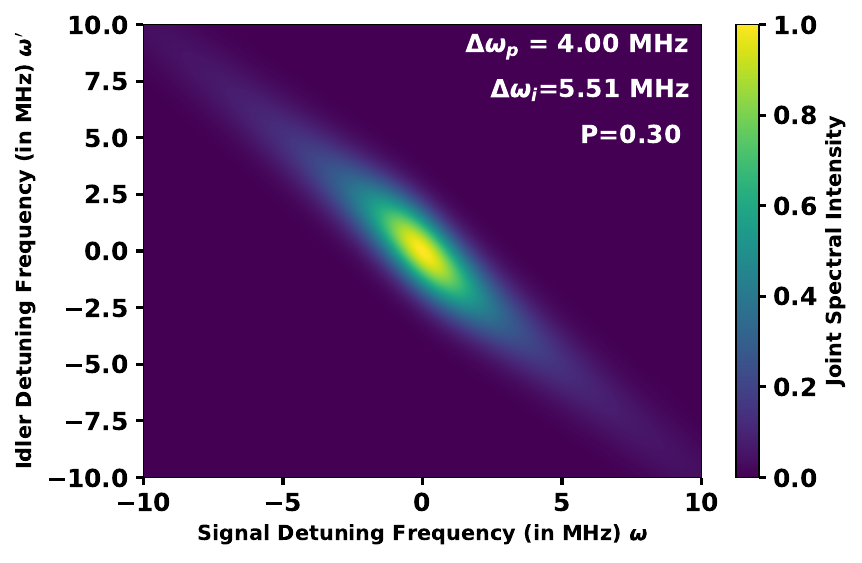}}


    \subfloat[]{\includegraphics[width=0.3\textwidth]{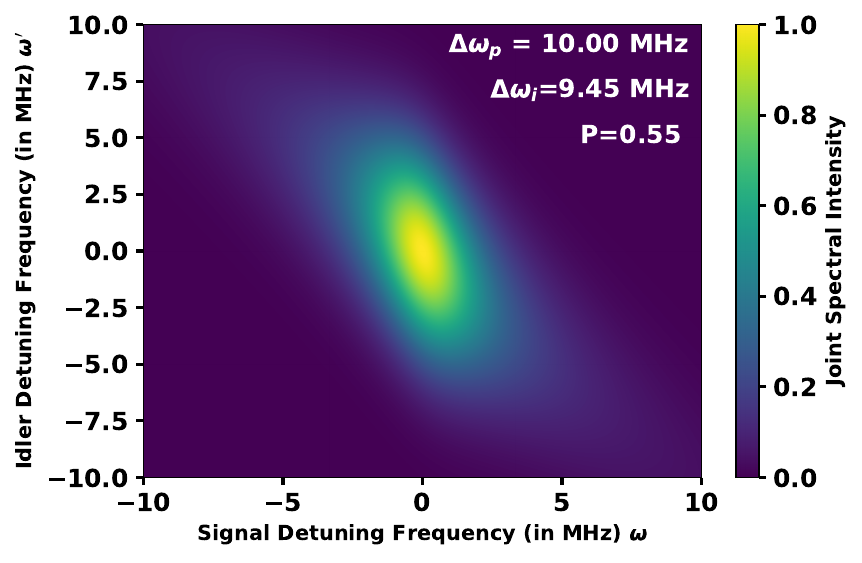}}
    \hfill
    \subfloat[]{\includegraphics[width=0.3\textwidth]{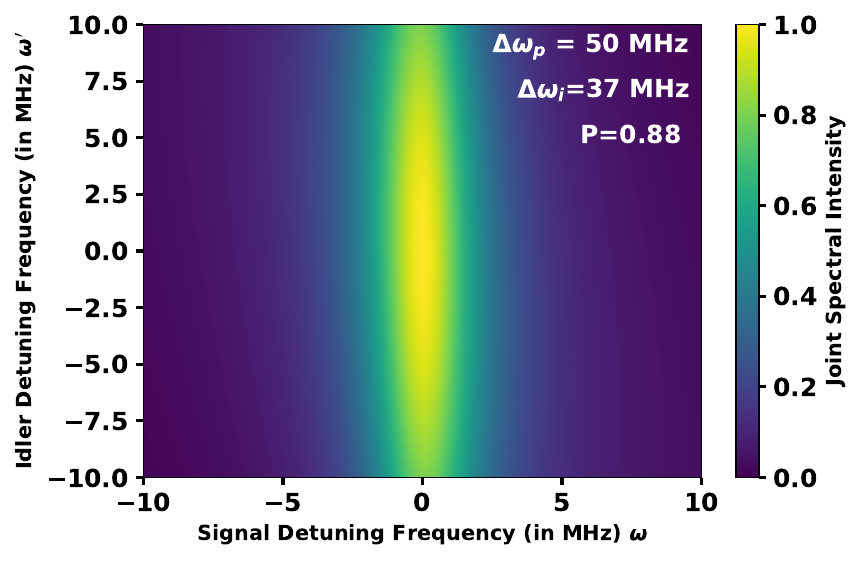}}
    \hfill
    \subfloat[]{\includegraphics[width=0.3\textwidth]{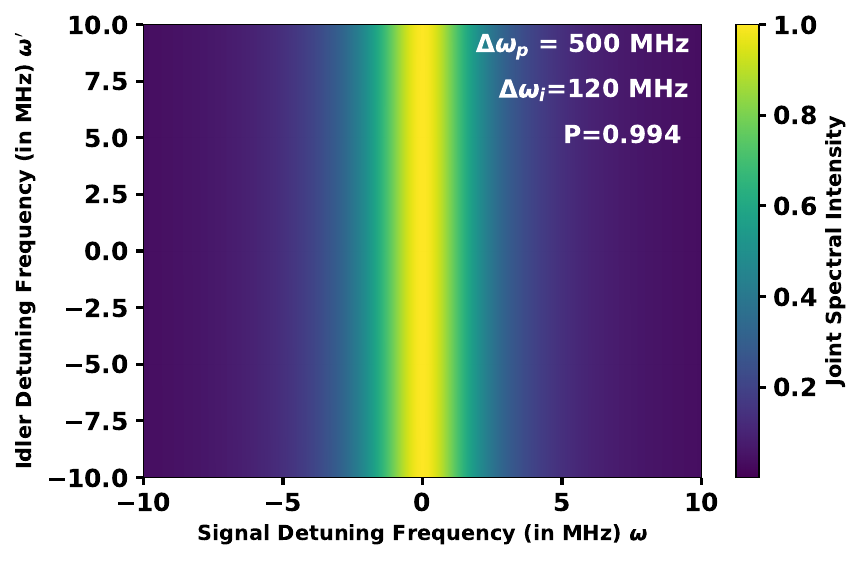}}

    \caption{\textbf{Normalized Joint Spectral Intensity with increasing pump bandwidth for a bare cavity width of 1~GHz and narrowed cavity width of 4~MHz. The signal bandwidth remains equal to the narrowed cavity width across all pump bandwidths. The pump bandwidth, $\Delta\omega_p$, idler bandwidth $\Delta\omega_i$, and purity, $P$, are noted on each plot.}}
    \label{fig:sixpanel-jsi}
\end{figure*}
 
\section{Broadband Pump} \label{sec:broadband}
Up to this point, we have assumed a continuous-wave, single frequency pump field. This will always generate spectrally entangled signal and idler photons, which typically cannot be used for entanglement swapping and similar quantum operations. Spectrally separable photons can be produced by using a pump field with bandwidth sufficiently large to erase the spectral correlations, typically achieved by pulsing the pump field appropriately. In a triply-resonant configuration (as we are assuming for the bare cavity case), this condition is challenging to meet as the pump field cannot be broader than the pump cavity mode, though it can be achieved if the pump's cavity mode can be engineered to be broader than the cavity modes for the signal and idler photons. In our proposed scheme, the pump cavity mode is naturally much broader than the narrowed signal and/or idler modes and thus the pump bandwidth can be much larger than the signal bandwidth.

We must now account for the non-stationary nature of the broadened pump and include its frequency dependence in the Fourier domain by replacing the scalar amplitude of the pump cavity mode, $\beta$, with a pump function, $\beta(\omega)$. For simplicity, we assume a Gaussian pump spectral function. We investigate the broadband pump case for the singly-filtered scheme to illustrate the effect on the bandwidth of the photon produced in the bare, unfiltered mode. We write down the signal and idler modes in the Fourier domain for a pump field with non-zero bandwidth in the singly-filtered regime:
\begin{widetext}
\begin{align}
 a(\omega)&=\frac{\frac{\sqrt{\kappa_a}}{n_g}}{\frac{\kappa_a}{2n_g}-i\omega}a_{\text{in}}(\omega) +\int_{-\infty}^\infty\frac{d\omega^\prime}{2\pi}v(\omega, \omega^\prime)b^\dagger_{\text{in}}(-\omega^\prime-2\delta_a)\\
 b^\dagger(-\omega-2\delta_a)&=\frac{\sqrt{\kappa_b}}{\frac{\kappa_b}{2}-i(\omega+\delta_b+\delta_a )}b^\dagger_{\text{in}}(-\omega-2\delta_a)+\sqrt{\frac{\kappa_a}{\kappa_b}}\int_{-\infty}^\infty\frac{d\omega^\prime}{2\pi}v^*(\omega^\prime, \omega)a_{\text{in}}(\omega^\prime).
\end{align}
\end{widetext}
Here, $v(\omega, \omega^\prime)$ is the joint spectral amplitude for the generated photon pairs. 

\begin{align}
v(\omega,\omega^\prime)=\frac{2ig\sqrt{\kappa_b}\frac{i\omega-\Delta}{\frac{\kappa_a}{2}+\frac{\kappa_{\text{abs}}}{2}}F(\omega)\beta(\omega-\omega^\prime)}{\left(\frac{\kappa_a}{2n_g} -i\omega\right)\left(\frac{\kappa_b}{2}-i(\omega^\prime+\delta_b+\delta_a)\right)}
\end{align}

\begin{figure}
    \centering
    \subfloat[]{
      \includegraphics[width=0.88\linewidth]{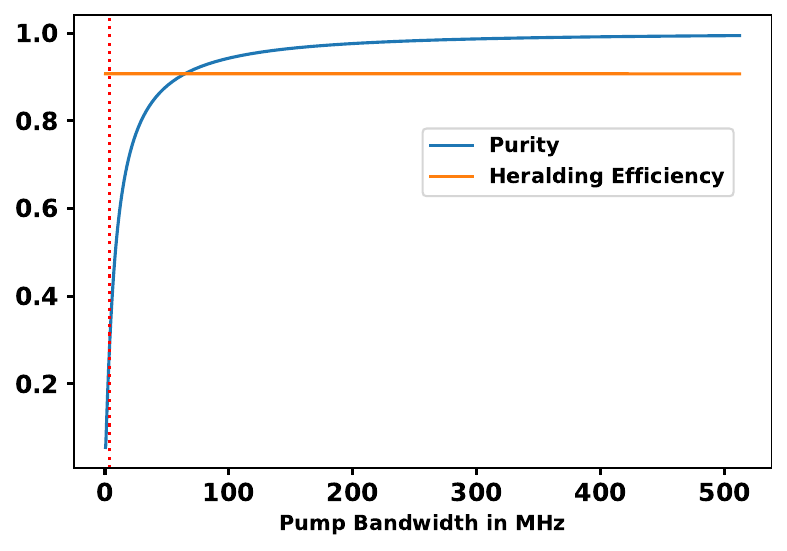}
      \label{fig:her_purity_1}
    }\\
    \subfloat[]{
      \includegraphics[width=0.88\linewidth]{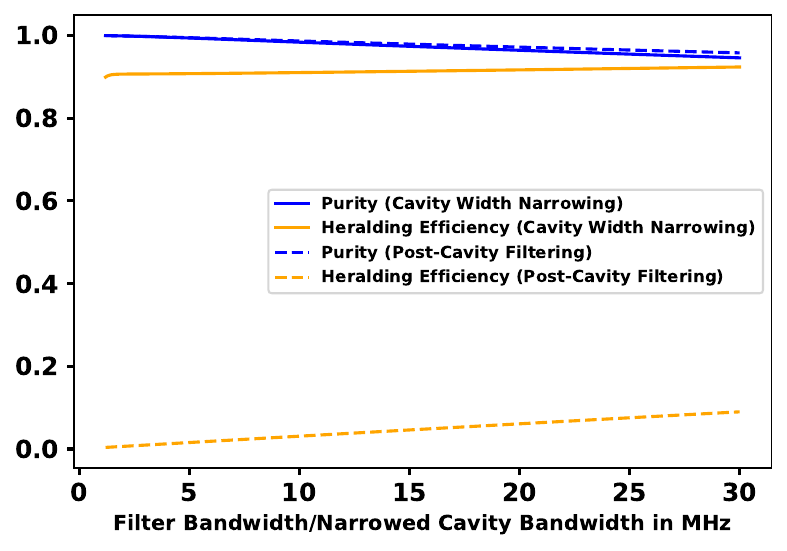}
      \label{fig:her_purity_2}
    }
     \caption{\textbf{(a) Purity and heralding efficiency as a function of pump bandwidth for a 1\,GHz wide bare cavity and 68\,MHz wide intra-cavity filter. Dotted vertical line denotes the narrowed cavity bandwidth. (b) Purity and heralding efficiency for intra-cavity filtering and post-cavity filtering for equivalent signal photon linewidths. The pump bandwidth is fixed at 600\,MHz }}
     \label{fig:heralding efficiency_purity}
\end{figure}

We numerically calculate the joint spectral intensity, $|v(\omega_s, \omega_i)|^2$ and show the results in Fig \ref{fig:sixpanel-jsi}. We also evaluate the purity and heralding efficiency as a function of the pump bandwidth. We consider a plausible experimental realization of an extended Lorentz filter-- a persistent spectral hole created in a dense, spectrally broad ensemble of coherent erbium ions doped in lithium niobate microrings. The spectral hole has a high degree of spectral configurability with the width and center of the spectral hole being determined by a resonant laser drive as seen in\,\cite{barya2025ultra, zhao2024cavity}. The center of the spectral hole can be tailored to overlap with a cavity mode. The cavity mode into which the other photon is emitted is assumed to be spectrally far removed from the inhomogeneously broadened ensemble and unaffected by the dopants. All the parameters in Fig \ref{fig:heralding efficiency_purity} are motivated by previous experimental realizations of SPDC in periodically poled lithium niobate microrings and spectral hole burning in erbium doped microrings. They have been listed in the supplemental material\,\cite{SuppMat}. We use the parameters $\kappa_{\text{abs}}=16~\text{GHz}$, $\Delta=68~\text{MHz}$, and $\kappa_a=\kappa_b=\kappa=1~\text{GHz}$ for the absorption due to the erbium dopants, the spectral hole width, and the bare cavity linewidth, respectively. This gives $n_g=250$ and a narrowed cavity width of 4~MHz. We note that $\Delta$ is widely tunable from $\lesssim10~\text{MHz}$ to $\gtrsim100~\text{MHz}$ in such a setup and we have chosen a mid-range value to illustrate the effect. We see that as the pump bandwidth exceeds the narrowed cavity width, the purity approaches unity while the heralding efficiency remains unchanged (see fig. \ref{fig:her_purity_1}). The heralding efficiency is determined solely by the escape efficiency i.e. the ratio of external coupling rate to total cavity decay rate, which always limits heralding efficiency for cavity enhanced photon pair production.  It is important to note that a narrow post-cavity filter can achieve high purity as well, but it is accompanied by a significant drop in heralding efficiency.  This is best demonstrated in Fig \ref{fig:her_purity_2} where we have compared  the performance of a post-cavity filter and an intr-cavity filter. The spectral hole width is varied such that the effective narrowed cavity width matches the bandwidth of the post-cavity filter whereas the pump bandwidth is kept fixed at a mid-range value of 600\,MHz. We see yet again that the heralding efficiency for the intra-cavity filter is nearly constant over the entire range. The heralding efficiency is not only lower for a post-cavity filter but it also drops significantly as filter bandwidth is reduced. 

\begin{figure*} 
    \centering

    \subfloat[]{%
        \includegraphics[width=0.32\textwidth]{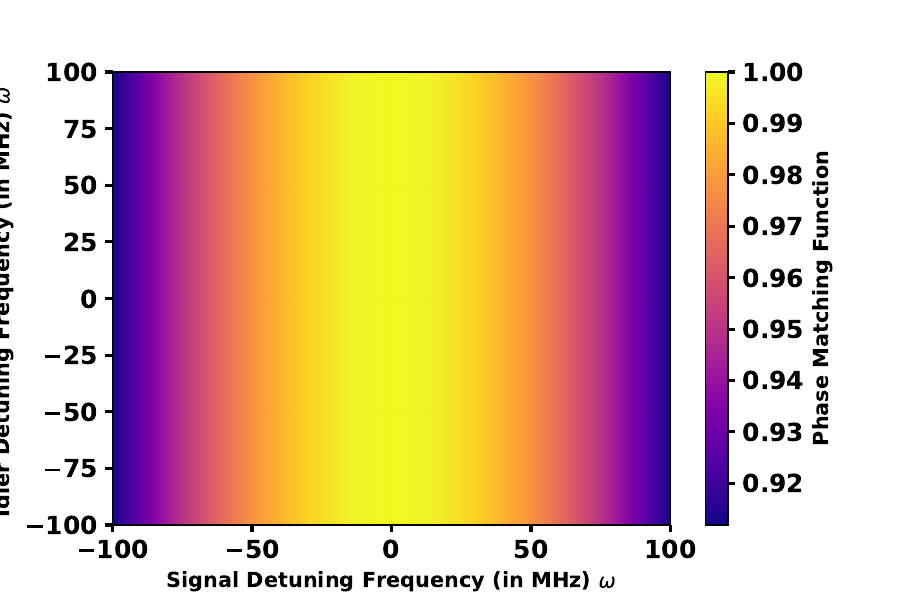}%
        \label{fig:panel1}%
    }\hfill
    \subfloat[]{%
        \includegraphics[width=0.32\textwidth]{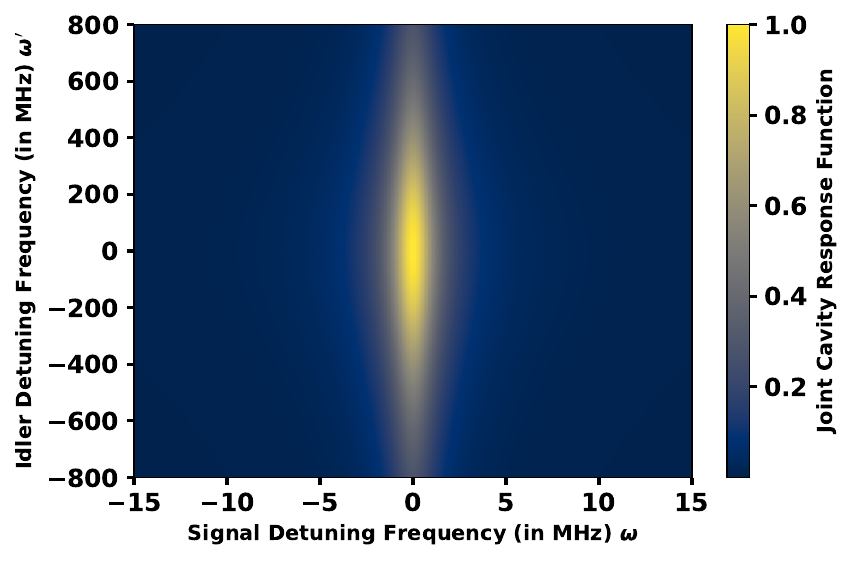}%
        \label{fig:panel2}%
    }\hfill
    \subfloat[]{%
        \includegraphics[width=0.32\textwidth]{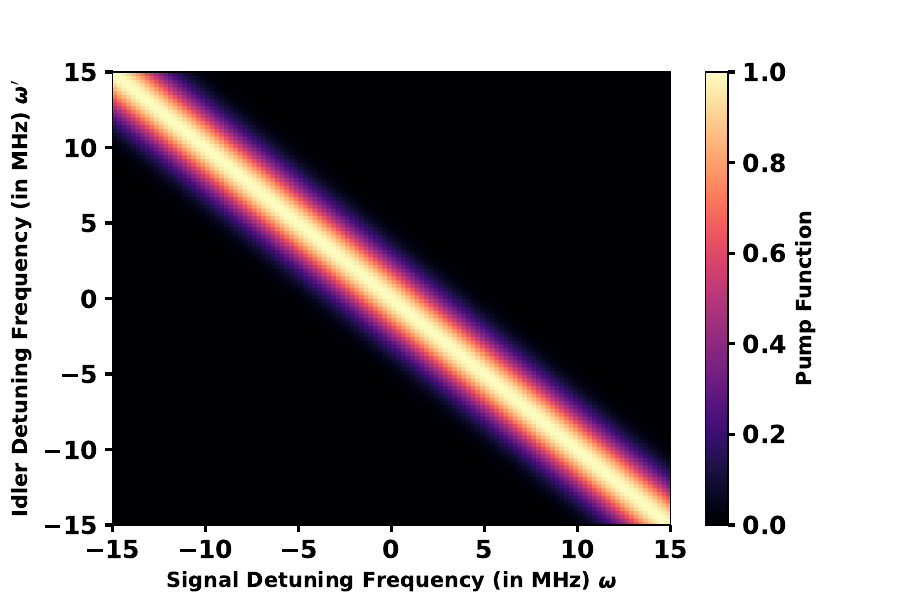}%
        \label{fig:panel3}%
    }

    \caption{ \textbf{(a) Phase Matching Function (b) Joint Cavity Response Function with idler cavity width of 1\,GHz and signal cavity width of 4\,MHz (c) Pumping function with pump width of 6MHz}}
    \label{fig:threepanel}
\end{figure*}

The trend of increasing spectral purity with pump bandwidth is best understood through the joint spectral intensity between the signal and the idler\,\cite{grice1998spectral,mauerer2009colors}. The signal and idler are spectrally uncorrelated when the joint spectral intensity has its major and minor axes parallel to the signal and idler frequency axes. Conversely, they are strongly correlated when the joint spectral intensity is diagonal. The shape of the joint spectral intensity is determined  by the compounded effect of the phase matching function $F(\omega)$, the pumping function $\left(\beta(\omega+\omega^\prime)\right)$ and the cavity response functions. The function with the narrowest spectral extent carves out the joint intensity and governs the tilt. 

The phase matching function in our scheme is effectively a function of solely the narrowed cavity mode detuning and thus extends vertically (See Fig \ref{fig:panel1}). For reasonable finesse values, the width of the phase matching function greatly exceeds that of the joint cavity response function. Thus, the phase-matching function exerts limited influence on the shape and tilt of the joint intensity. Due to the asymmetric widths of the cavity modes, the joint cavity response function is elongated vertically while the pumping function is diagonal. The narrower of the two determines the tilt, so a sufficiently broad pump leads to spectrally pure states across a wide range of parameter regimes. 

In a triply resonant cavity configuration, the pump bandwidth cannot be larger than the width of the pump cavity mode, which is typically comparable to the bare, unfiltered signal or idler cavity width for an integrated photonic cavity. When the cavity in question is a microring, the triply resonant configuration is the only possible implementation: if the pump drive is not resonant with a cavity mode, light will not couple evanescently from the waveguide into the microring due to destructive interference. This is in sharp contrast to bulk resonators where one can tailor the transmission characteristics of the input and output mirrors to create any of the three possible resonant configurations-- singly, doubly and triply resonant. For example, in the singly resonant configuration, the coupling mirrors are reflective at the signal wavelength and highly transmissive for the pump and idler wavelengths. Thus, the pump and idler photons escape the cavity after a single pass through the SPDC crystal. This allows for highly broadband pumps, which, as we have seen, is essential for achieving large purity. Such wavelength dependent cavity width engineering is extremely challenging for microrings. We believe our intra-cavity filter scheme is a promising approach for engineering effective cavity widths and increasing of configurability and tunability of on-chip photon pair production.  

\section{Discussion}
We have demonstrated that our intra-cavity filtering scheme provides a pathway to effectively increasing the length of a cavity without increasing the physical length, whose benefits for photon pair production are clear. This is crucial for integrated photonics applications where the size of devices and propagation losses limit the practically achievable on-chip cavity sizes. We have shown, using experimentally plausible parameters, that it is possible to generate narrowband photon pairs with high single photon purity and without sacrificing the heralding efficiency. 

For numerical simulations, we have focused on a specific implementation of a spectral hole-based dissipative bandpass filter, but our treatment is broad and can be extended to other implementations. Bandpass filter widths narrower than the cavity widths of integrated photonic resonators can most likely be achieved using a range of coherent emitters, from atoms to ions to color center defects\,\cite{lukin2025mesoscopic, aghaeimeibodi2018integration, gritsch2022narrow}. Our work goes past work to date where photonic structures have been used to enhance the capability of coherent quantum emitters\,\cite{gonzalez2024light, thompson2013coupling, dibos2018atomic, zhou2024trapped}. We are exploring new regimes where the quantum capabilities of the emitters are being used to enhance the capabilities of photonic structures.

\section{Acknowledgments}
We thank Priyash Barya, Paul Kwiat, and Brian DeMarco for useful discussions. This material is based upon work supported by the National Science Foundation under grant no. 2143172.

\FloatBarrier
\bibliography{main_bib_hyperref}
\bibliographystyle{apsrev4-2}

\end{document}


\maketitle
The supplemental material is structured in the following manner. Section A elaborates on the narrowing of the cavity width and the formalism we adopt to describe it. We focus on the doubly filtered degenerate case but the same formalism applies  to the doubly and singly filtered non-degenerate case as well. Section B and C focuses on deriving the general form of the expressions for the performance metrics listed in Table 1 and 2 in the main text for the continuously pumped degenerate and non-degenerate case respectively. Finally, in section D, we derive the expressions for the spectral purity and heralding efficiency for a broadband pump. 
\subsection{Cavity Width Narrowing }
As described in the main text, we introduce an extended Lorentz type bandpass dissipative filter in the cavity with full width half-maximum of $\Delta$ and centered at the cavity resonance frequency. We choose a Lorentz type filter since it is analytically tractable but the general results are applicable to filters with other profiles as well. In the frame of reference of the bare cavity resonance frequency, $\omega_a$,  we have
\begin{align}
     \dot{a}=& -2ig\beta a^\dagger -\frac{\kappa_a}{2}a-\frac{\kappa_{\text{abs}}}{2}a +\frac{\kappa_{\text{abs}}}{2}\frac{\Delta}{2}\int_{0}^td\tau\, e^{-\frac{\Delta}{2}(t-\tau)} a(\tau)+\sqrt{{\kappa_a}} a_{\text{in}}
\end{align}
The equation assumes a simpler form in the Fourier domain:
\begin{align}
-i\omega\,a(\omega)-&i\omega\left(\frac{\Delta-i\omega}{\kappa_a+\kappa_{\text{abs}}}\right)a(\omega)=-\frac{\kappa_a}{2}\frac{\Delta}{\kappa_a+\kappa_{\text{abs}}} a(\omega)-2ig\beta\frac{\Delta-2i\omega}{\kappa_a+\kappa_{\text{abs}}}a^\dagger(-\omega-2\delta_a)+\sqrt{\kappa_a}\frac{\Delta-2i\omega}{\kappa_a+\kappa_{\text{abs}}}a_{\text{in}}(\omega)\\
-i\omega\,\tilde{a}(\omega)&= \frac{\kappa_a}{2}\frac{\Delta}{\kappa_a+\kappa_{\text{abs}}} \tilde{a}(\omega)-2ig\beta\frac{\Delta-2i\omega}{\kappa_a+\kappa_{\text{abs}}}\tilde{a}^\dagger(-\omega-2\delta_a)+\sqrt{\kappa_a}\frac{\Delta-2i\omega}{\kappa_a+\kappa_{\text{abs}}}a_{\text{in}}(\omega)\\
-i\omega\,\tilde{a}(\omega)&= \frac{\kappa_a}{2}\frac{\Delta}{\kappa_a+\kappa_{\text{abs}}} \tilde{a}(\omega)-2ig\beta\frac{\Delta}{\kappa_a+\kappa_{\text{abs}}}\tilde{a}^\dagger(-\omega-2\delta_a)+\sqrt{\kappa_a}\frac{\Delta}{\kappa_a+\kappa_{\text{abs}}}a_{\text{in}}(\omega)
\end{align}
The first of these equations is an exact Fourier transform, albeit with some rearrangement of terms and coefficients. In the second equation, we drop the term  $-i\omega\left(\frac{\Delta-i\omega}{\kappa_a+\kappa_{\text{abs}}}\right)a(\omega)$. This is valid only when we are focusing on the slow dynamics induced by the filter and corresponds to an implicit Fano-Feschbach projection. We have added the tilde over the operator $(\tilde{a})$ to indicate the same. In the final equation,  we have made two approximations : $2ig\beta\frac{\Delta-i\omega}{\kappa_a+\kappa_{\text{abs}}}\tilde{a}^\dagger(-\omega-2\delta_a)\approx 2ig\beta\frac{\Delta}{\kappa_a+\kappa_{\text{abs}}}\tilde{a}^\dagger(-\omega-2\delta_a)$ and $\sqrt{\kappa_a}\frac{\Delta-2i\omega}{\kappa_a+\kappa_{\text{abs}}}a_{\text{in}}(\omega)\approx \sqrt{\kappa_a}\frac{\Delta}{\kappa_a+\kappa_{\text{abs}}}a_{\text{in}}(\omega)$. The latter of the two is justified by the fact that $a_{\text{in}}(\omega)$ is the vacuum mode and its time derivative is zero. When the pumping rate $\beta$ is below the optical parametric oscillator threshold, as is assumed to be the case throughout this work, the conjugate mode, $a(\omega+2\delta_a)$ is coupled mainly to the vacuum mode and the frequency term is dropped for the same reason. We have confirmed the validity of these two approximations through numerical simulations. The final results for all metrics remain unchanged with or without these last set of approximations. Just as in the main text, we identify the quantity $\frac{\kappa_a+\kappa_{\text{abs}}}{\Delta}$ to be the group index at the center of the filter and replace future instances in the supplementary information with $n_g$.
\subsection{Continuous Pump Degenerate SPDC} 
The effective Langevin equations for the degenerate signal and idler modes in the Fourier domain are as follows:
\begin{align}
-i\omega\, \tilde{a}(\omega)&=-\frac{\kappa_a}{2}\frac{1}{n_g}\tilde{a}(\omega)-2ig\beta\frac{1}{n_g}\tilde{a}^\dagger(-\omega-2\delta_a)+\sqrt{\kappa_a}\frac{1}{n_g}a_{\text{in}}(\omega)\\
-i(\omega+2\delta_a)\, \tilde{a}^\dagger(-\omega-2\delta_a)&=-\frac{\kappa_a}{2}\frac{1}{n_g}a^\dagger(-\omega-2\delta_a)+2ig\beta\frac{1}{n_g}a(\omega)+\sqrt{\kappa_a}\frac{1}{n_g}\tilde{a}^\dagger_{\text{in}}(-\omega-2\delta_a)
\end{align}
The above pair of equations can be solved to give:
\begin{align}
\tilde{a}(\omega)=\frac{\sqrt{\kappa_a}}{\frac{\kappa_a}{2n_g}-i\omega}\frac{1}{n_g}a_{\text{in}}(\omega)-\frac{2ig\beta\sqrt{\kappa_a}}{\left(\frac{\kappa_a}{2n_g}-i(\omega+2\delta_a)\right)\left(\frac{\kappa_a}{2n_g}-i\omega\right)}\frac{1}{n^2_g}a^\dagger_{\text{in}}(-\omega-2\delta_a)
\end{align}
Note that we have assumed the pumping rate to be below the optical parametric oscillator threshold and used approximations to that effect in arriving at the above expression. The intrinsic spectral density of the generated photons is as follows:
\begin{align}
    &S_a(\omega, \omega^\prime)=\kappa_a\langle \tilde{a}^\dagger(\omega)\tilde{a}(\omega^\prime)\rangle\nonumber\\
    &=\kappa_a\frac{4g^2|\beta|^2\kappa_a}{\left[\frac{\kappa_a}{2n_g}+i\left(\omega+2\delta_a\right)\right]\left[\frac{\kappa_a}{2n_g}-i\left(\omega^\prime+2\delta_a\right)\right]\left[\frac{\kappa_a}{2n_g}+i\omega\right]\left[\frac{\kappa_a}{2n_g}-i\omega^\prime\right]}\frac{1}{n_g^2}\langle a_{\text{in}}(-\omega-2\delta_a)a^\dagger_{\text{in}}(-\omega^\prime-2\delta_a)\rangle\\
    &=\kappa_a\frac{4g^2|\beta|^2\kappa_a}{\left[\frac{\kappa_a}{2n_g}+i\left(\omega+2\delta_a\right)\right]\left[\frac{\kappa_a}{2n_g}-i\left(\omega^\prime+2\delta_a\right)\right]\left[\frac{\kappa_a}{2n_g}+i\omega\right]\left[\frac{\kappa_a}{2n_g}-i\omega^\prime\right]}\frac{1}{n_g^2}2\pi\delta(\omega-\omega^\prime)
\end{align}
Thus, spontaneous parametric downconversion arises from parametric amplification of vacuum noise . We qualify the photon spectral density as ``intrinsic" to suggest that we are considering the photons escaping the cavity through all leakage channels. The output photon spectral density $\left(S_{a, \text{out}}(\omega, \omega^\prime)\right)$ is related to the intrinsic density as:
\begin{align}
S_{a,\text{out}}(\omega, \omega^\prime) =\frac{{\kappa_{a}}_{\text{ext}}}{\kappa_a}S_a(\omega, \omega^\prime)
\end{align}
where ${\kappa_a}_{\text{ext}}$ is the external coupling coefficient. The intrinsic photon generation rate is simply the two-dimensional Fourier transform of the spectral density.
\begin{align}
R=&\frac{1}{2}\int\int_{-\infty}^\infty\frac{d\omega}{2\pi}\frac{d\omega^\prime}{2\pi}S_a(\omega, \omega^\prime)e^{i(\omega-\omega^\prime)t}\\   
=&\frac{1}{2}\int_{-\infty}^\infty\frac{d\omega}{2\pi}\kappa_a\frac{4g^2|\beta|^2\kappa_a}{\left[\left(\frac{\kappa_a}{2n_g}\right)^2+\left(\omega+2\delta_a\right)^2\right]\left[\left(\frac{\kappa_a}{2n_g}\right)^2+\omega^2\right]}\frac{1}{n_g^2}\\
=&\frac{g^2|\beta|^2\kappa_a}{\delta_a^2+\left(\frac{\kappa_a}{2n_g}\right)^2}\frac{1}{n_g^3}
\end{align}
We arrive at the final expression using contour integration. In order to investigate the temporal correlation between the signal and the idler, we compute the second-order correlation function:
\begin{align}
      G^{(2)}(\tau)=&\langle \tilde{a}_{\text{out}}^\dagger(t)\tilde{a}_{\text{out}}^\dagger(t+\tau)\tilde{a}_{\text{out}}(t+\tau)\tilde{a}_{\text{out}}(t)\rangle
\end{align}
The output field is determined by the input-output relations:
\begin{align}
    \tilde{a}_{\text{out}}&={a_{\text{in}}}_{\text{ext}}-\sqrt{{\kappa_a}_{\text{ext}}}\tilde{a}\nonumber\\
    a_{\text{in}}&=\sqrt{\frac{{\kappa_a}_{\text{ext}}}{\kappa_a}}{a_{\text{in}}}_{\text{ext}}+\sqrt{\frac{{\kappa_a}_{\text{int}}}{\kappa_a}}{a_{\text{in}}}_{\text{int}}
\end{align}
The input vacuum field, $a_{\text{in}}$ that appears in the dynamical equations is implicitly understood to comprise of the vacuum mode coupling to the microring cavity via the waveguide, ${a_{\text{in}}}_{\text{ext}}$ and the modes coupling to the cavity via the scattering channels, ${a_{\text{in}}}_{\text{int}}$. The rate at which the cavity couples to the scattering channels determines the internal loss rate, ${\kappa_a}_{\text{int}}$.  
\begin{align}
      G^{(2)}(\tau)=&4g^2\left|\frac{\beta}{n_g^2}\right|^2{\kappa_a}_{\text{ext}}^2\left|\int_{-\infty}^\infty\frac{d\omega}{2\pi}\frac{e^{i\omega\tau}}{\left[\frac{\kappa_a}{2n_g}+i(\omega+2\delta_a)\right]\left[\frac{\kappa_a}{2n_g}-i\omega\right]}\right|^2\\
      =&{\kappa_a}_{\text{ext}}^2\frac{g^2\left|\frac{\beta}{n_g^2}\right|^2}{\delta_a^2+\left(\frac{\kappa_a}{2n_g}\right)^2}e^{-\frac{\kappa_a}{n_g}|\tau|}
\end{align}
The heralding efficiency can be determined from the photon pair generation rate and the correlation function
\begin{align}
    \eta^h=&\frac{\int_{-T/2}^{T/2}d\tau\,G^{(2)}(\tau)}{\frac{2{\kappa_a}_{\text{ext}}}{\kappa_a}R}\nonumber\\
    =&\frac{{\kappa_a}_{\text{ext}}}{\kappa_a}(1-e^{-\frac{\kappa_a}{n_g}\frac{T}{2}})
\end{align}
\subsection{Continuous-wave pumped singly-filtered SPDC}
We consider spontaneous parametric downconversion into a pair of spectrally disparate cavity modes, of which only one of them has been subjected to the action of an extended intra-cavity filter. The Langevin equations in the Fourier domain and in the rotating frame of the narrowed cavity mode, $\tilde{a}$ are as follows:
\begin{align}
-i\omega\tilde{a}(\omega)=-\frac{\kappa_a}{2}\frac{1}{n_g}\tilde{a}(\omega)-2ig\beta F(\omega) \frac{1}{n_g}\tilde{a}^\dagger(-\omega-2\delta_a)
+\sqrt{{\kappa_a}}\frac{1}{n_g}a_{\text{in}}(\omega)\\
-i\omega b^\dagger(-\omega-2\delta_a)= i(\delta_b+\delta_a) b^\dagger (-\omega-2\delta_a)
-\frac{\kappa_b}{2} b^\dagger(-\omega-2\delta_a)+2ig\beta^*F^*(\omega)\tilde{a}(\omega)+\sqrt{{\kappa_b}}b_{\text{in}}(\omega)	 
\end{align}
Here, $F(\omega)$ is the phase matching function with the following form:
\begin{align}
	 F(\omega)&=\text{sinc}\left(\frac{\omega\tau_{\text{rt}}}{2}\right)\\
	\tau_{\text{rt}}&=\frac{nL}{v_c}(n_g-1)\approx \frac{nL}{v_c}n_g  \hspace{1cm} n_g \gg 1  
\end{align}

Solving for the pair of conjugate modes, we obtain:
\begin{align}
	\tilde{a}(\omega)&=\frac{\sqrt{\kappa_a}}{\frac{\kappa_a}{2n_g}-i\omega}\frac{1}{n_g}a_{\text{in}}(\omega)-\frac{2ig\beta\sqrt{\kappa_b}}{\left(\frac{\kappa_a}{2n_g}-i\omega\right)\left(\frac{\kappa_b}{2}-i\omega-i(\delta_b+\delta_a)\right)}\frac{F(\omega)}{n_g}b^\dagger_{\text{in}}(-\omega-2\delta_a)\\
	b^\dagger(-\omega-2\delta_a)&=\frac{\sqrt{\kappa_b}}{\frac{\kappa_b}{2}-i\omega-i(\delta_b+\delta_a)}b^\dagger_{\text{in}}(-\omega-2\delta_a)+\frac{2ig\beta^*\sqrt{\kappa_a}}{\left(\frac{\kappa_a}{2n_g}-i\omega\right)\left(\frac{\kappa_b}{2}-i\omega-i(\delta_b+\delta_a)\right)}\frac{F^*(\omega)}{n_g}a_{\text{in}}(\omega)
\end{align}
We have once again assumed the pumping rate to be far below the optical parametric oscillator threshold and made the appropriate approximations. We can now determine the intrinsic spectral density of each cavity mode. For the narrowed cavity mode, we have: 
\begin{align}
	&S_a(\omega, \omega^\prime)=\kappa_a\langle \tilde{a}^\dagger(\omega)\tilde{a}(\omega^\prime)\rangle\nonumber\\	
	&=\kappa_a\frac{4g^2|\beta|^2\kappa_b F^*(\omega)F(\omega^\prime)}{\left[\frac{\kappa_a}{2n_g}+i\omega\right]\left[\frac{\kappa_b}{2}+i\omega+i(\delta_b+\delta_a)\right]\left[\frac{\kappa_a}{2n_g}-i\omega^\prime\right]\left[\frac{\kappa_b}{2}-i\omega^\prime-i(\delta_b+\delta_a)\right]}\frac{1}{n_g^2}\langle b_{\text{in}}(-\omega-2\delta_a)b^\dagger_{\text{in}}(-\omega^\prime-2\delta_a)\rangle\\
	&=\kappa_a\frac{4g^2|\beta|^2\kappa_b F^*(\omega)F(\omega^\prime)}{\left[\frac{\kappa_a}{2n_g}+i\omega\right]\left[\frac{\kappa_b}{2}+i\omega+i(\delta_b+\delta_a)\right]\left[\frac{\kappa_a}{2n_g}-i\omega^\prime\right]\left[\frac{\kappa_b}{2}-i\omega^\prime-i(\delta_b+\delta_a)\right]}\frac{1}{n_g^2}2\pi\delta(\omega-\omega^\prime)
\end{align}
We can similarly solve for the spectral density of the photons generated in cavity mode $ b$  i.e. $S_b(\omega, \omega^\prime)$. We find that the spectral density is identical upto a frequency detuning. Thus, when the pumping rate is below the optical parametric threshold, the spectral density is simply a product of two Lorentzians with widths $\frac{\kappa_a}{n_g}$ and $\kappa_b$. The bandwidth $(\Delta\omega)$ or the full width half maximum for such a product is a well-known result and has the following form:
\begin{align}
	\Delta\omega&=\sqrt{\frac{\sqrt{\left(\frac{\kappa_a}{n_g}\right)^4+6\left(\frac{\kappa_a}{n_g}\kappa_b\right)^2+\kappa_b^4}-\left[\left(\frac{\kappa_a}{n_g}\right)^2+\kappa_b^2\right]}{2}}\nonumber\\
    &\approx \frac{\kappa_a}{n_g}
\end{align}
In the second line, we have performed a binomial approximation on the inner square root since $\frac{\kappa_a}{n_g}\ll \kappa_b$. This demonstrates that when there is a large disparity between the cavity widths, the overall bandwidth is  determined by the narrowest width. The intrinsic photon generation rate can be derived from the spectral density with the result:
\begin{align}
R=&\int\int_{-\infty}^\infty\frac{d\omega}{2\pi}\frac{d\omega^\prime}{2\pi}S_a(\omega, \omega^\prime)e^{i(\omega-\omega^\prime)t}\\ 
=&\int_{-\infty}^\infty \frac{d\omega}{2\pi}\kappa_a\frac{4g^2|\beta|^2\kappa_b |F(\omega)|^2}{\left[\left(\frac{\kappa_a}{2n_g}\right)^2+\omega^2\right]\left[\frac{\kappa_c^2}{4}+(\omega+\delta_b+\delta_a)^2\right]}\label{eq: R_nondeg_1}\\
=&\frac{4g^2|\beta|^2\kappa_b\kappa_a}{\tau_{\text{rt}}\left(\frac{\kappa_a}{2}\right)^2\left(\frac{\kappa_b}{2}\right)^2} + \frac{4g^2|\beta|^2\kappa_b\kappa_a}{\tau_{\text{rt}}^2n_g^2\left[\left(\frac{\kappa_b}{2}\right)^2-\left(\frac{\kappa_a}{2n_g}\right)^2\right]}\left[\frac{1-e^{-\frac{\kappa_b}{2}\tau_{\text{rt}}}}{\left(\frac{\kappa_b}{2}\right)^3}-\frac{1-e^{-\frac{\kappa_a}{2n_g}\tau_{\text{rt}}}}{\left(\frac{\kappa_a}{2n_g}\right)^3}\right]\label{eq: R_nondeg_2}
\end{align}
Eq \eqref{eq: R_nondeg_2} was obtained from \eqref{eq: R_nondeg_1} using contour integration with a contour in the upper half complex plane. We note that the quantities in the exponential: $\frac{\kappa_b\tau_{\text{rt}}}{2}$ and $\frac{\kappa_a\tau_{\text{rt}}}{2n_g}$ can be rewritten succinctly in terms of finesse and group index such that:
\begin{align}
\frac{\kappa_b\tau_{\text{rt}}}{2}&=\frac{\pi n_g}{\mathcal{F}} \nonumber\\
\frac{\kappa_a\tau_{\text{rt}}}{2n_g}&=\frac{\pi}{\mathcal{F}}
\end{align}
This of course assumes that each of the two cavity modes have nearly identical bare cavity widths and finesse. Thus, there are two parameter regimes that naturally arise: $\mathcal{F}\gg n_g\gg 1$ and $n_g\gg\mathcal{F}\gg 1$, both of which are experimentally accessible. In the limit, $n_g\gg \mathcal{F}\gg 1$, we  approximate the term, $e^{-\frac{\kappa_b\tau_{\text{rt}}}{2}}=e^{-\frac{\pi n_g}{\mathcal{F}}}$ to zero whereas for the limit, $\mathcal{F}\gg n_g\gg 1$, we perform a Taylor series expansion upto the first order (or the second order if the first order expansion term cancels out). For the second exponential term, $e^{-\frac{\kappa_a\tau_{\text{rt}}}{2n_g}}=e^{-\frac{\pi}{\mathcal{F}}}$, we  perform a Taylor series expansion in both regimes. In either limit, the intrinsic photon generation rate converges to the same value reported in Table 2 of the main text.  

We now focus on the second order correlation function, starting with the correlation function conditioned on mode $b$ as the idler mode. We denote it as $G^{(2)}_{a/b}(\tau)$ and it has the following form: 
\begin{align}
    G^{(2)}_{a/b}(\tau)&=\langle b_{\text{out}}^\dagger(t)\tilde{a}_{\text{out}}^\dagger(t+\tau)\tilde{a}_{\text{out}}(t+\tau)b_{\text{out}}(t)\rangle\nonumber\\
    &=\frac{{\kappa_a}_{\text{ext}}}{\kappa_a}\frac{{\kappa_b}_{\text{ext}}}{\kappa_b}R^2+4g^2\left|\frac{\beta}{n_g}\right|^2{\kappa_a}_{\text{ext}}{\kappa_b}_{\text{ext}}\left|\int_{-\infty}^\infty\frac{d\omega}{2\pi}\frac{F(\omega)e^{-i\omega\tau}}{\left(\frac{\kappa_a}{2n_g}-i\omega\right)\left(\frac{\kappa_b}{2}+i\omega\right)}\right|^2
\end{align}
We have set the detunings to zero here. The first term arises from the accidental coincidences. The second term assumes different forms in different temporal regimes relative to the round-trip time $\tau_{\text{rt}}$. Physically, this is unsurprising since the slow photons in the filtered mode are still circulating in the cavity when the faster photons occupying the bare cavity mode escape. Algebraically, different  temporal regimes lead to different choice of contours when the above integral is analytically extended to the complex plane. When the time difference exceeds half the round-trip time i.e. when $\tau> \frac{\tau_{\text{rt}}}{2}$, a semicircular contour in the upper half complex plane encloses a simple pole at $i\frac{\kappa_a}{2n_g}$ and results in:
\begin{align}
    G^{(2)}_{a/b}(\tau)&=\frac{{\kappa_a}_{\text{ext}}}{\kappa_a}\frac{{\kappa_b}_{\text{ext}}}{\kappa_b}R^2+4g^2\left|\frac{\beta}{n_g}\right|^2{\kappa_a}_{\text{ext}}{\kappa_b}_{\text{ext}}\frac{e^{-\frac{\kappa_a}{n_g}\tau}}{\left(\frac{\kappa_b}{2}+\frac{\kappa_a}{2n_g}\right)^2}\text{sinc}^2\left(i\frac{\kappa_a\tau_{\text{rt}}}{4ng}\right)
\end{align}
We now determine the correlation function , $G^{(2)}_{a/b}(\tau)$ for the situation $\tau<\frac{\tau_{\text{rt}}}{2}$. An appropriate contour for this regime is a semicircle lying in the lower half complex plane which encloses a simple pole at $-i\frac{\kappa_b}{2}$.
\begin{align}
  G^{(2)}_{a/b}(\tau)&=\frac{{\kappa_a}_{\text{ext}}}{\kappa_a}\frac{{\kappa_b}_{\text{ext}}}{\kappa_b}R^2+4g^2\left|\frac{\beta}{n_g}\right|^2{\kappa_a}_{\text{ext}}{\kappa_b}_{\text{ext}}\frac{e^{\kappa_b\tau}}{\left(\frac{\kappa_b}{2}+\frac{\kappa_a}{2n_g}\right)^2}\text{sinc}^2\left(i\frac{\kappa_b\tau_{\text{rt}}}{4}\right)
\end{align}
Finally, the temporal regime that remains is $-\frac{\tau_{\text{rt}}}{2}<\tau<\frac{\tau_{\text{rt}}}{2}$. In this regime, the two exponential terms of the sinc function decay to zero at infinity in different halves of the complex plane and thus, need to be evaluated individually. The sinc function is analytic at zero despite an apparent singularity. However, that is no longer the case when two exponential terms are evaluated separately and the pole at zero needs to be accounted for by deforming the contour. 
\begin{align}
  G^{(2)}_{a/b}(\tau)&=\frac{{\kappa_a}_{\text{ext}}}{\kappa_a}\frac{{\kappa_b}_{\text{ext}}}{\kappa_b}R^2+\frac{4g^2\left|\frac{\beta}{n_g}\right|^2{\kappa_a}_{\text{ext}}{\kappa_b}_{\text{ext}}}{\left(\frac{\kappa_b}{2}+\frac{\kappa_a}{2n_g}\right)^2}\left[\frac{e^{-\frac{\kappa_a}{2n_g}\left(\tau+\frac{\tau_{\text{rt}}}{2}\right)}}{\frac{\kappa_a}{2n_g}\tau_{\text{rt}}}-\frac{1}{\frac{\kappa_a}{2n_g}\tau_{\text{rt}}}+\frac{e^{\frac{\kappa_b}{2}\left(\tau-\frac{\tau_{\text{rt}}}{2}\right)}}{\frac{\kappa_b}{2}\tau_{\text{rt}}}-\frac{1}{\frac{\kappa_b}{2}\tau_{\text{rt}}}\right]^2
\end{align}
Note that in previous analytical treatments of cavity enhanced SPDC with a continuous pump, the correlation function in this intermediate temporal regime has either been disregarded or incorrectly computed leading to a discontinuous and unphysical function\,\cite{scholz2009analytical,slattery2019background}. This error goes unnoticed in most experimental implementations since in a typical cavity enhanced SPDC experiment, the difference in the round trip time between the signal and the idler is picoseconds, shorter than all other timescales and beyond the detector resolution. The experimentally determined correlation function smooths the apparent erroneous discontinuity around $\tau=0$. However, in our case, the round trip time difference can be several nanoseconds, well within the resolution limits of modern detectors.  

We may similarly evaluate the correlation function conditioned on mode $\tilde{a}$. 
\begin{align}
    G^{(2)}_{b/a}(\tau)=&\langle \tilde{a}^\dagger(t)b^\dagger(t+\tau)b(t+\tau)\tilde{a}(t)\rangle\nonumber\\
      G^{(2)}_{b/a}(\tau)=&\frac{{\kappa_a}_{\text{ext}}}{\kappa_a}\frac{{\kappa_b}_{\text{ext}}}{\kappa_b}R^2+4g^2|\beta|^2{\kappa_a}_{\text{ext}}{\kappa_b}_{\text{ext}}\left|\int_{-\infty}^\infty\frac{d\omega}{2\pi}\frac{F(\omega)e^{i\omega\tau}}{\left(\frac{\kappa_a}{2n_g}-i\omega\right)\left(\frac{\kappa_b}{2}+i\omega\right)}\right|^2\nonumber\\
      G^{(2)}_{b/a}(\tau)=&G^{(2)}_{a/b}(-\tau)
\end{align}
We see that for a continuous pump, switching the ordering of the two modes is equivalent to flipping the sign of the time difference. With the knowledge of the correlation function and pair generation rate, it is possible to determine the heralding efficiency but as one might anticipate, the general analytic expressions are prohibitively complicated and provide little insight. In the limits: $\mathcal{F}\gg n_g\gg 1$ and $n_g\gg\mathcal{F}\gg 1$, one can obtain approximate expressions for the heralding efficiency.
\begin{table*}[t]
\centering
\begin{tabular}{cc}
\toprule
  \textbf{$T < \tau_{\text{rt}}$} & \textbf{$T > \tau_{\text{rt}}$} \\
\midrule
$\displaystyle
\frac{\kappa_{\text{ext}}}{\kappa}
\frac{\kappa}{n_g}
\frac{T(T^2+3\tau_{\text{rt}}^2)}
{12\tau_{\text{rt}}^2}
$
&
$\displaystyle
\frac{\kappa_{\text{ext}}}{\kappa}
\left(
1-e^{-\frac{\kappa}{n_g}\frac{T}{2}}
\right)
-
\frac{\kappa_{\text{ext}}}{\kappa}
\frac{\kappa\tau_{\text{rt}}}{6n_g}
$
\\
\bottomrule
\end{tabular}
\caption{\textbf{Heralding efficiency, $\eta^h$, for an ideal box detection window of time $T$ in the regime where the finesse and filter-induced group index are large: $\mathcal{F}, n_g \gg 1$.}}
\label{table:her_eff}
\end{table*}

\subsection{Broadband Pump}
When the pump is broadband, amplitude of the pump drive is non-stationary and  becomes frequency dependent in the Fourier domain. The signal and idler modes in the Fourier domain in the below-OPO regime are now modified as follows:
\begin{align}
 a(\omega)&=\frac{\frac{\sqrt{\kappa_a}}{n_g}}{\frac{\kappa_a}{2n_g}-i\omega}a_{\text{in}}(\omega) +\int_{-\infty}^\infty\frac{d\omega^\prime}{2\pi}v(\omega, \omega^\prime)b^\dagger_{\text{in}}(-\omega^\prime-2\delta_a)\\
 b^\dagger(-\omega-2\delta_a)&=\frac{\sqrt{\kappa_b}}{\frac{\kappa_b}{2}-i(\omega+\delta_b+\delta_a )}b^\dagger_{\text{in}}(-\omega-2\delta_a)+\sqrt{\frac{\kappa_a}{\kappa_b}}\int_{-\infty}^\infty\frac{d\omega^\prime}{2\pi}v^*(\omega^\prime, \omega)a_{\text{in}}(\omega^\prime)
\end{align}
where $v(\omega, \omega^\prime)$ is the joint spectral amplitude and the quantity $|v(\omega, \omega^\prime)|^2$ is referred to as the joint spectral intensity.
\begin{align}
    v(\omega,\omega^\prime)=\frac{2ig\sqrt{\kappa_b}\frac{i\omega-\Delta}{\frac{\kappa_a}{2}+\frac{\kappa_{\text{abs}}}{2}}F(\omega)\beta(\omega-\omega^\prime)}{\left(\frac{\kappa_a}{2n_g} -i\omega\right)\left(\frac{\kappa_b}{2}-i(\omega^\prime+\delta_b+\delta_a)\right)}
\end{align}
 Note that while plotting the joint spectral intensity in the main text, we have performed a change of variable: $\omega^\prime\rightarrow -\omega^\prime $. It has been shown that the spectral purity ($P$) is related to the experimentally measurable normalized autocorrelation function $\tilde{g}(0)$ with the tilde denoting that the measurement scheme employed (typically a photon counter) has a detection window that exceeds the correlation time between the conjugate modes with the relation being $P=\tilde{g}^{(2)}(0)-1$\,\cite{mauerer2009colors}. This is implicitly assumed and is typically experimentally satisfied with broadband SPDC schemes with no cavity enhancement. However, with narrowband SPDC generation, the bandwidth of the square law detector needs careful consideration. 
\begin{align}
    \tilde{g}^{(2)}(0)=1+\frac{\int_{-\infty}^\infty \frac{d\omega_1d\omega_2d\omega^\prime d\omega^{\prime\prime}}{(2\pi)^4}v^*(\omega_1, \omega^{\prime\prime})v(\omega_2, \omega^{\prime\prime})v^*(\omega_2, \omega^\prime)v(\omega_1, \omega^\prime)}{\int_{-\infty}^\infty\left|\frac{d\omega_1d\omega^\prime}{(2\pi)^2}|v(\omega_1,\omega^\prime)|^2\right|^2}
\end{align}
Numerically computing a quadruple integral is challenging and if the integrand is oscillatory, prohibitively so. We instead discretize the joint spectral amplitude function, $v(\omega, \omega^\prime)$ onto a finely spaced, uniform grid $(\omega_i, \omega^\prime_j)$ producing the matrix $V_{ij}$. Following the discretization, the integrals can be replaced by a matrix trace operation, which are significantly easier to compute. 
\begin{align}
    \tilde{g}^{(2)}(0)=1+\frac{\text{Tr}\left(V^\dagger VV^\dagger V\right)}{
\left[\text{Tr}(V^\dagger V)\right]^2}
\end{align}
Post-cavity filtering schemes lead to improvements in spectral purity at the cost of heralding efficiency. We evaluate the heralding efficiency to show that our extended intra-cavity filtering does not suffer from the same tradeoff. We consider the case where the narrowed cavity mode  $\tilde{a}$ is the signal mode and $b$ is the idler mode. 
\begin{align}
    \eta^h=\frac{{\kappa_a}_{\text{ext}}}{2\kappa_a}\frac{\left|\int_{-\infty}^\infty\frac{d\omega d\omega^\prime}{(2\pi)^2}\frac{\frac{\kappa_a}{2n_g}+i\omega}{\frac{\kappa_a}{2n_g}-i\omega}v(\omega, \omega^\prime)\right|^2}{\int_{-\infty}^\infty\frac{d\omega d\omega^\prime}{(2\pi)^2}|v(\omega, \omega^\prime)|^2}+\frac{{\kappa_c}_{\text{ext}}}{2\kappa_c}\frac{\left|\int_{-\infty}^\infty\frac{d\omega d\omega^\prime}{(2\pi)^2}\frac{\kappa_c+i\omega}{\kappa_c-i\omega}v(\omega, \omega^\prime)\right|^2}{\int_{-\infty}^\infty\frac{d\omega d\omega^\prime}{(2\pi)^2}|v(\omega, \omega^\prime)|^2}
\end{align}
We resort to the same discretization as earlier to numerically compute the efficiency. In Table 2, we have listed the parameter values which we used for all the plots in the main text. These values have been drawn from our previous work and previous works on SPDC generation in lithium niobate rings by other authors. 
\begin{table*}[t]
\centering
\begin{tabular}{lc}
\toprule
Pump Cavity Width ($\kappa_b$) 
& $2\pi\,1\,\text{GHz}$ \\

Bare Cavity Width ($\kappa_a$) 
& $2\pi\,1\,\text{GHz}$ \\

Bare Cavity Width ($\kappa_c$) 
& $2\pi\,1\,\text{GHz}$ \\

External Cavity Coupling Rate (${\kappa_c}_{\text{ext}}$)
& $2\pi\,900\,\text{MHz}$\\

External Cavity Coupling Rate (${\kappa_a}_{\text{ext}}$)
& $2\pi\,900\,\text{MHz}$\\

Absorption Loss Rate ($\kappa_{\text{abs}}$)
& $2\pi\,16\,\text{GHz}$ \\

Finesse ($\mathcal{F}$)
& $150$ \\

Spectral Hole Width ($\Delta$)
& $2\pi\,68\,\text{MHz}$
 \\

Non-linearity Coupling Rate ($g$)
& $2\pi\,2\,\text{MHz}$\\

Pumping Rate ($\epsilon_p$)
& $2\pi\, 3\,\text{GHz}$\\
\midrule\midrule\\

Group Index at Hole Center ($n_g=\frac{\kappa_a+\kappa_{\text{abs}}}{\Delta}$)
& $250$\\
Pump Cavity Amplitude ($\beta=\frac{\epsilon_p}{\frac{\kappa_b}{2}+i\delta_b}$)
& $20$\\
Roundtrip Time Difference ($\tau_{\text{rt}}=\frac{2\pi n_g}{\kappa_a\mathcal{F}}$)
& $2.8\,\text{ns}$\\
\bottomrule
\end{tabular}
\caption{\textbf{Parameter values for plots in the main text unless otherwise stated. All cavity detunings were set to zero. The second half lists the derived parameters.}}
\end{table*}

\bibliography{main_bib_hyperref}
\bibliographystyle{ieeetr}